\documentclass[usenatbib]{mn2e}
\usepackage{graphicx}

\def\apj{\rm ApJ}
\def\apjl{\rm ApJL}

\def\aj{\rm AJ}
\def\mnras{\rm MNRAS}
\def\nat{\rm Nature}

\def\pasp{\rm PASP}
\def\aap{\rm AAP}
\def\araa{\rm ARA\&A}

\def\prd{\rm PRD}

\def\gax{\mathrel{\raise.3ex\hbox{$>$}\mkern-14mu\lower0.6ex\hbox{$\sim$}}}
\def\lax{\mathrel{\raise.3ex\hbox{$<$}\mkern-14mu\lower0.6ex\hbox{$\sim$}}}
\def\gtorder{\mathrel{\raise.3ex\hbox{$>$}\mkern-14mu
             \lower0.6ex\hbox{$\sim$}}}
\def\ltorder{\mathrel{\raise.3ex\hbox{$<$}\mkern-14mu
             \lower0.6ex\hbox{$\sim$}}}

\begin{document}

\title [TDE Demographics]
   {Tidal Disruption Event (TDE) Demographics }

\author[C.~S. Kochanek]{ C.~S. Kochanek$^{1,2}$\\
  $^{1}$ Department of Astronomy, The Ohio State University, 140 West 18th Avenue, Columbus OH 43210 \\
  $^{2}$ Center for Cosmology and AstroParticle Physics, The Ohio State University,
    191 W. Woodruff Avenue, Columbus OH 43210 
   }

\maketitle

\begin{abstract}
We survey the properties of stars destroyed in TDEs as
a function of BH mass, stellar mass and evolutionary state, star formation
history and redshift.  For $M_{BH}{\ltorder}10^7 M_\odot$,
the typical TDE is due to a $M_*{\sim}0.3M_\odot$ M-dwarf, although
the mass function is relatively flat for $M_*{\ltorder}M_\odot$. The contribution from
older main sequence stars and sub-giants is small but not negligible.  From
$M_{BH}{\simeq}10^{7.5}$-$10^{8.5M_\odot}$, the balance rapidly shifts to
higher mass stars and a larger contribution from evolved stars, and is ultimately
dominated by evolved stars at higher BH masses.  The star formation history
has little effect until the rates are dominated by evolved stars.  TDE rates should
decline very rapidly towards higher redshifts.  The volumetric rate of TDEs
is very high because the BH mass function diverges for low masses.  However,
any emission mechanism which is largely Eddington-limited for low BH
masses suppresses this divergence in any observed sample and
leads to TDE samples dominated by
$M_{BH}{\simeq}10^{6.0}$-$10^{7.5} M_\odot$ BHs with roughly
Eddington peak accretion rates.
The typical fall back time is relatively long, with 16\%
having $t_{fb}<10^{-1}$~years (37 days), and 84\% having longer time scales.
Many residual rate discrepancies can be explained if surveys are biased
against TDEs with these longer $t_{fb}$, which seems very plausible if
$t_{fb}$ has any relation to the transient rise time.  For almost any BH
mass function, systematic searches for fainter, faster time scale TDEs in
smaller galaxies, and longer time scale TDEs in more massive galaxies
are likely to be rewarded.
\end{abstract}


\section{Introduction}
\label{sec:introduction}

Tidal Disruption Events (TDEs) occur when a star passes sufficiently
close to a supermassive black hole for the tidal fields to destroy (or severely maim)
the star 
(\citealt{Hills1975}, \citealt{Lacy1982}, \citealt{Carter1983}, \citealt{Rees1988}, \citealt{Evans1989}).
In the last $\sim 10$ years, significant numbers of TDEs have begun to be
discovered (see, e.g., the reviews by \citealt{Gezari2012} and
\citealt{Komossa2015}).  
The first candidates were mostly found as either X-ray or UV flares 
in archival data  (see the summary in \citealt{Komossa2015}).
More recently, large scale transient surveys like ASAS-SN
(\citealt{Shappee2014}), PTF (\citealt{Law2009}) and
Pan-STARRS (\citealt{Kaiser2002}) have found increasing
numbers of TDEs in real time, allowing more detailed photometric
and spectroscopic follow up studies (e.g., \citealt{Holoien2014},
\citealt{Holoien2016}, \citealt{vanvelzen2011}, \citealt{Gezari2012}, \citealt{Arcavi2014}).

Demographic studies of TDEs 
have largely focused on the dynamical problem of understanding
the rate at which stars can be placed into low angular momentum orbits
that will pass sufficiently close to the black hole to be disrupted
(e.g., \citealt{Lightman1977}, \citealt{Cohn1978}, 
\citealt{Magorrian1999}, \citealt{Wang2004}, \citealt{Merritt2005}, \citealt{Brockamp2011}, \citealt{Vasiliev2013}).  
These studies generally considered only main sequence stars with a common
mass and structure, although
\cite{Magorrian1999} discusses the effect of a mass function on
the rates, while \cite{Syer1999} and \cite{Macleod2012} considered
the relative rates for main sequence and evolved stars at a fixed
mass but not general populations.  
\cite{Strubbe2009} used an evolving black hole mass function
and a model for the observational properties of TDEs to
estimate rates for a range of observational surveys.
\cite{Mageshwaran2015} and
\cite{Stone2016} considered a population of main sequence stars
mainly following the \cite{Kroupa2001} initial mass function (IMF) 
truncated at $M_\odot$ to mimic an old stellar population but
otherwise ignored star formation histories and stellar evolution.
\cite{Mageshwaran2015} focused on absolute rate estimates in
various scenarios, while \cite{Stone2016} also examine the 
distribution of event properties.  Broadly speaking, most theoretical studies 
predict TDE rates of order $10^{-4}$~yr$^{-1}$~gal$^{-1}$ for $M_{BH} \ltorder 10^{7.5}M_\odot$, while many
observational rate estimates are closer to $10^{-5}$~yr$^{-1}$~gal$^{-1}$
(e.g. \citealt{Donley2002}, \citealt{Gezari2008}, \citealt{vanvelzen2014}).
\cite{Holoien2016}, however, found a somewhat higher rate 
in the All-Sky Automated Survey for Supernovae (ASAS-SN).

There are extensive numerical studies of the hydrodynamics
of TDEs (e.g., \citealt{Evans1989}, \citealt{Lodato2009}, \citealt{Macleod2012}, \citealt{Dai2013}, 
\citealt{Guillochon2013}, \citealt{Hayasaki2013}, \citealt{Guillochon2014},
\citealt{Shiokawa2015}, \citealt{Bonnerot2016}, \citealt{Sadowski2015}) and semi-analytic models of their observational
properties (e.g., \citealt{Rees1988}, \citealt{Cannizzo1990},
\citealt{Kochanek1994}, 
\citealt{Loeb1997}, \citealt{Strubbe2009}, \citealt{Syer1999},
\citealt{Kasen2010}, \citealt{Lodato2011}, 
\citealt{Strubbe2011}, \citealt{Stone2013},
\citealt{Miller2015}, \citealt{Piran2015}, \citealt{Strubbe2015}, \citealt{Metzger2015}).  
However, it is fair to say that these
studies have yet to converge on a predictive model for TDE properties.
The fundamental difficulty is that TDEs are a three dimensional
radiation hydrodynamics problem.  Simulations are still
challenged by the large range of spatial scales and do not yet 
include the effects of radiation, while semi-analytic models
are not well-suited for transients with necessarily complex
spatial structures.  

Our goal in this paper is to examine the demographics of TDEs.
Given the lack of any reliable predictive
model for observables, we focus on a simple, generic model
for selection effects that can be observationally calibrated.
In our
models, we include not only an initial mass function for stars,
but also star formation histories and complete models of
stellar evolution.  We use a mass function for the black holes
as well as estimates of its evolution with redshift.  
In \S2
we describe our model for the stellar populations, disruption
rates and the black hole mass function.  In \S3 we survey the expected demographics of TDEs
as a function of black hole mass, stellar mass, stellar
evolutionary state, and redshift both globally and for
a specific observational case.  We will discuss
the implications and directions for further inquiry in \S4.

\section{Model Description}

In this section we outline the model we will use for this study.
We start with the criteria for disrupting (or stripping sufficient
mass to cause a flare) a star of mass $M_*=M_{*\odot}M_\odot$ and 
radius $R_*=R_{*\odot} R_\odot$.  Then we introduce a mass
function for the stars, star formation histories and a model
for stellar evolution.  Next we estimate the rates of disruptions
for a bulge with velocity dispersion $\sigma=200\sigma_{200}$~km/s 
containing a black hole of 
mass $M_{BH}=10^7 M_{BH7} M_\odot$, and discuss black hole
mass functions.  Finally, we examine several 
physical properties of disruptions and introduce a simple
selection effects model.   

We assume that an event occurs when a star approaches
closer to the black hole than
\begin{equation}
      R_T = R_* \left( \eta^2 { M_{BH} \over M_* }\right)^{1/3}
       \label{eqn:disruption}
\end{equation}
where $\eta \simeq 1$.  If the pericentric radius $R_p$ is larger than
the Schwarzschild radius, $R_S=2 G M_{BH}/c^2$, but smaller than $R_T$, then we assume
there is some form of TDE.  If it is smaller than the Schwarzschild radius, we assume 
the star falls into the black hole and is absorbed without a luminous 
transient.  Arguably, we might instead use the radius of the 
last stable orbit.  Depending on
the structure of the star and the exact pericenter, the 
star may be fully destroyed or only stripped of all or
part of its envelope (e.g. \citealt{Macleod2012}).  In addition to the mass of the black
hole, the detailed limits ($\eta$ etc.) depend weakly
on the structure of the star (e.g. \citealt{Macleod2012}, \citealt{Guillochon2013})
and the properties of the black hole (e.g. \citealt{Kesden2012}).

We assume a \cite{Kroupa2001} initial mass function (IMF) 
extending from $0.08 M_\odot$ to $100 M_\odot$.
This makes the IMF a broken power law, $(dn/dM_*)_{IMF} \propto (M_*/0.5 M_\odot)^{-\alpha}$
with $\alpha = 1.3$ for $M_*<0.5 M_\odot$ and $\alpha=2.3$
for $M_*>0.5 M_\odot$. The complete \cite{Kroupa2001}
IMF breaks to a still shallower $\alpha=0.3$ power-law
at $0.08M_\odot$ and extends down to $0.01 M_\odot$,
but we ignore this extension to brown dwarfs.  Observational
selection effects will disfavor finding such low mass TDEs
in any case.
The  mass function at any given time, $dn/dM_*$, is not the IMF, due
to the combined effects of stellar evolution and the star formation
history.  Where stellar mass functions have previously been
used in TDE rates studies (\citealt{Magorrian1999}, \citealt{Mageshwaran2015},
\citealt{Stone2016}), they assume an old stellar population
by simply truncating the IMF at a 
maximum mass of $M_\odot$.  While this is a reasonable
model for the present day main sequence population of an
early type galaxy, it is not a good characterization of the
central regions of the Milky Way. \cite{Pfuhl2011}, for example,
find that the Galactic center population is mostly old
(80\% formed 5-10~Gyr ago), but the remainder is
in a very young population (20\% formed in the last
$\sim 0.1$~Gyr). If the typical TDE occurs in a galaxy
with $M_{BH} \sim 10^6$ to $10^7M_\odot$, the mixed
stellar population we see in the Milky Way may be more
representative than a purely old population.  It is also natural
to include broader models of star formation histories
since we include stellar evolution and will explore
the evolution of TDE rates with redshift,

We considered two basic star formation histories, a 1~Gyr
burst and continuous star formation, with the star formation
rate constant during the star forming period. We examine the
resulting TDE rates and properties at ages of
$1$, $3$ and $10$~Gyr.  The two histories are the same
at an age of 1~Gyr, so there are really 5 distinct cases.
For example, if the life time of a star is $t_*(M_*)$, then
\begin{equation}
   { dn \over dM_*} 
       \propto \left( { dn \over dM_*} \right)_{IMF}
            \hbox{min} \left( t, t_*(M_*) \right)
       \label{eqn:mf}
\end{equation}
is the mass function at time $t$ for a constant star formation rate and
ignoring mass loss.  It follows the IMF until the mass where
$t = t_*(M_*)$ and is then cut off because only star formation
in the last $t_*(M_*)$ contributes to the mass function.
The burst model simply requires more accounting to include the
effects of the cutoff in star formation.  We define the mass
function so that it
is normalized, $\int dM_* dn/dM_* \equiv 1$, and it is useful
to define the mean stellar mass $\langle M_* \rangle$ and
the mean square mass $\langle M_*^2 \rangle$

We use the older Padua stellar isochrones of \cite{Marigo2008}
because they include the thermally pulsating AGB (TP-AGB) phase
of stellar evolution.  We considered only the Solar metallicity
models.  The \cite{Marigo2008} tracks
start at $M_* = 0.15 M_\odot$, so we extended them
down to $M_* = 0.08 M_\odot$ by logarithmically extrapolating
physical quantities (luminosity, temperature) with mass.  The 
exact details are not critical -- the primary goal is simply 
to better estimate the absolute numbers of low mass stars.
We also tracked the population of stellar remnants.  We used
the initial-to-final white dwarf mass relation $M_{WD}=0.109M_{ZAMS}+0.394 M_\odot$ 
from \cite{Kalirai2008} for $M_* < 8M_\odot$, neutron 
star masses of $1.4M_\odot$ from $8M_\odot$ to $21.4M_\odot$, and 
black hole masses of $7M_\odot$ for higher masses.  
These choices are broadly consistent with the observed properties of supernova
progenitors (\citealt{Smartt2009}), typical black hole 
masses (e.g. \citealt{Ozel2010}, \citealt{Kreidberg2012}, \citealt{Kochanek2015})
and estimates of the fraction of core collapses leading to black holes 
($\sim 25\%$, \citealt{Kochanek2015}).  The detailed distribution
of neutron star and black hole outcomes with stellar mass is an
open question (see \citealt{Kochanek2015}), but unimportant here.

We track the evolutionary state of the stars in five bins using the
tags supplied by the \cite{Marigo2008} isochrones.  We track
main sequence stars (MS, up to the turn off tag TO), sub-giant
stars (from TO to the base of the red giant branch, RGBb),
red giants (RGB stars, from RGBb to helium ignition, BHeb),
horizontal branch stars (HB stars, from helium ignition BHeb to core
helium exhaustion, EHeb), asymptotic giant branch stars
(AGB, from core helium exhaustion to carbon ignition Cb),
and stars after carbon ignition. In practice, this latter
phase makes a negligible contribution and can simply be
ignored.  We will always use the description of the phases 
as MS, sub-giant, RGB, HB and AGB.  The categories are not
fully correct for the most massive stars (and the full
sequence of tags is not present), but they are appropriate for 
the intermediate mass stars that dominate the disruption rates 
of evolved stars.  Using a power law fit to the main sequence
turn off age, we also track the elapsed main sequence lifetime
fraction $f_{MS}$ of each star.

We start from the TDE rate estimate of \cite{Wang2004} modified to
include the effects of a mass function (see \citealt{Magorrian1999}),
\begin{equation}
   {d r \over dM_*} \simeq { 7.28 \langle M_*^2 \rangle^{3/8} R_*^{1/4} \eta^{1/6} \sigma^{7/2} 
       \over G^{5/4} M_{BH}^{11/12} M_*^{1/12} \langle M_*\rangle } { dn \over dM_*}. 
    \label{eqn:rate}
\end{equation}
The disruption rate for an individual star scales as 
$r \propto R_*^{1/4} M_*^{-1/12} \propto R_T^{1/4}$ as noted by \cite{Macleod2012}.
In this expression, we have neglected an additional term of the form
$(\ln \Lambda/\ln B)^{3/4} \sim 1$ where 
$\Lambda = 0.4 M_{BH}/M_0$, $B= r_h/4 R_T$ and $r_h=G M_{BH}/\sigma^2$
for simplicity. 
Since the $\eta^{1/6}$ dependence on the dimensionless factor
setting the disruption boundary from Equation~\ref{eqn:disruption}
is very weak, we simply set $\eta \equiv 1$.
This rate estimate also neglects any role of stellar collisions in
suppressing disruptions of giant stars (see \citealt{Macleod2012}).  
\cite{Wang2012} combine Equation~\ref{eqn:rate} with the 
$M_{BH}$-$\sigma$ relation
\begin{equation}
       M_{BH} \simeq 1.5 \times 10^8 \sigma_{200}^{4.65} M_\odot
      \label{eqn:msig}
\end{equation} 
to yield a rate scaling as $r \propto \sigma^{-3/4}$ or $r \propto M_{BH}^{-1/6}$
with black hole mass.

Using Solar values for all the stellar variables in Equations~\ref{eqn:rate} and
\ref{eqn:msig}, the
absolute scale of the rate is of the form $dr/dM_* = r_0 M_{BH7}^\alpha dn/dM_*$
with $r_0 = 4.7 \times 10^{-4}$/year and $\alpha=-0.164$.  For comparison,
\cite{Wang2004} ultimately adopt $r_0 = 3.7 \times 10^{-4}$/year and $\alpha=-0.25$ 
based on numerical fits to their results for individual galaxies.
\cite{Stone2016}, based on similar models of a larger number of 
galaxies and averaging over a stellar mass function, find $r_0 = 7.4 \times 10^{-5}$/year
and $\alpha=-0.404$. Using a completely different approach, \cite{Brockamp2011}
find $r_0=8.3\times 10^{-5}$/year and $\alpha=+0.446$ or $r_0 = 1.0 \times 10^{-4}$/year
and $\alpha=+0.353$ depending on their choice of $M_{BH}$-$\sigma$ relations.
There are then further systematic uncertainties coming from the treatment of
the angular structure of the stellar core (e.g. \citealt{Magorrian1999}, 
\citealt{Vasiliev2013}) and the role of binary black holes (e.g. \citealt{Merritt2005},
\citealt{Chen2008}, \citealt{Li2015}).  As a compromise over these various
results, we adopt a rate model of
\begin{equation}
   {d r \over dM_* } \simeq r_0 { \langle M_{*\odot}^2 \rangle^{3/8} R_{*\odot}^{1/4} \over 
        M_{BH7}^{1/4} M_{*\odot}^{1/12} \langle M_{*\odot} \rangle }  { dn \over dM_*}. 
    \label{eqn:rateuse}
\end{equation}
with $r_0 = 10^{-4}$/year and $\alpha=-1/4$, where the $*\odot$ subscript indicates that the quantity
is in Solar units.  We also neglect the process of a star ``evolving'' into its 
loss cone due to its increasing radius as it ascends the giant branch (\citealt{Syer1999})
since it is sub-dominant (\citealt{Magorrian1999}).  The effects of changing $r_0$
on our results are trivial, and we will explore the consequences of changing 
the dependence on the black hole mass ($\alpha$) below.

Equations~\ref{eqn:rate} and \ref{eqn:rateuse} depend on the mean stellar mass $\langle M_* \rangle$
and the mean square mass $\langle M_*^2 \rangle$.
The $\langle M_*\rangle$ term represents the change in the rate with the number
of stars at fixed total mass. The rate increases if the mass is divided over
larger numbers of stars.   The $\langle M_*^2 \rangle$ term is due to the 
dependence of the orbit diffusion rates on stellar mass -- the higher the
mean square stellar mass at fixed total mass, the faster the diffusion times, 
because the gravitational potential is becoming more ``granular'' (\citealt{Magorrian1999}).  
If we examine just the $\langle M_*^2 \rangle^{3/8}/\langle M_* \rangle$ factor
in Equation~\ref{eqn:rate}, the inclusion of a mass function (excluding remnants)
increases the rates by roughly a factor of $1.7$ and depends little on the age
or star formation history cases we consider.  This is consistent with \cite{Stone2016}, although
they also note that the factor increases significantly for very young ($\sim 100$~Myr) stellar
populations.  If we include stellar remnants in $\langle M_*^2 \rangle$ 
(they should not be included in $\langle M_* \rangle$ since the remnants make
a negligible contribution to disruption rates due to their high densities),
the rate increase is modestly higher, at a factor of roughly $2.0$.  \cite{Stone2016}, using theoretical
models of the black hole mass function by \cite{Belczynski2010}, found modestly
larger effect, and the differences likely lie in our using a lower, observationally
driven, choice for the typical black hole mass.  Since the effects of remnants
in our models are so small (10-20\% effects) we neglect them for simplicity. 

\begin{figure}
\centerline{\includegraphics[width=3.5in]{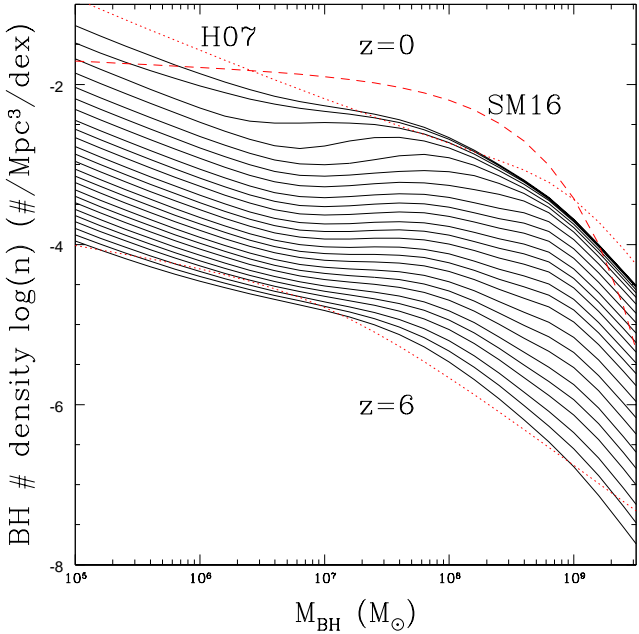}}
\caption{
  The evolution of the \protect\cite{Shankar2009} black hole mass function,
  $dn/d\log M_{BH}$, defined by the number of black holes per (base 10) logarithmic mass
  interval from the present (top) to $z=6$ (bottom) in steps of $\Delta z \simeq 0.25$. 
  The \protect\cite{Hopkins2007} (H07) mass function used by \protect\cite{Strubbe2009} is shown by
  red dotted lines at $z=0$ and $6$, and the local mass function used by \protect\cite{Stone2016}
  (SM16) is shown by the red dashed line.
  }
\label{fig:bhmf}
\end{figure}

To examine the overall rates of TDEs we need the black hole
mass function $n(M_{BH})$ as a function of redshift.  We
consider the models of \cite{Hopkins2007} and \cite{Shankar2009},
focusing on the more recent \cite{Shankar2009} models.  Both
models are based on using quasar luminosity functions and
estimates of merger rates to model the growth of black
holes, constrained by the requirement of matching local
estimates of the black hole mass function.  The \cite{Shankar2009},
mass function is defined per (base 10) logarithmic mass interval,
$n(M_{BH})=dn/d\log M_{BH}$, over the mass range $5.0 < \log M_{BH}/M_\odot < 9.6$  
and the redshift range $0 < z < 6$ as shown in Figure~\ref{fig:bhmf}.
The number of lower mass black holes is divergent ($\sim M_{BH}^{-3/2}$ for
low $M_{BH}$), although the
total mass in black holes is convergent.  This means that the 
observed rate of TDEs from low mass black holes is controlled by selection effects.
\cite{Strubbe2009} used the \cite{Hopkins2007} models and these
are very similar, as illustrated by their structure at $z=0$ and
$6$ in Figure~\ref{fig:bhmf}.  \cite{Mageshwaran2015}
used an evolving quasar luminosity function rescaled by a duty
cycle estimate, while \cite{Stone2016} start from a 
local galaxy luminosity function and then populate the
galaxies with black holes using the \cite{Mcconnell2013}
bulge/black hole mass correlation combined with a
bulge mass-dependent black hole occupation fraction.\footnote{In
Equation~31 of \cite{Stone2016}, the differential should be
$dM_{BH}$ rather than $d\ln M_{BH}$ (Stone, private communication).} 
This mass function, which has a shallower divergence $\sim M_{BH}^{-1.07}$
for low black hole masses, is also shown in Figure~\ref{fig:bhmf}.  
As emphasized by \cite{Stone2016}, the degree to which the divergence
of the number density for small $M_{BH}$ is real controls the absolute
volumetric rate of TDEs.

Some properties of TDEs may depend on the orbital pericentric radius $R_p$ 
relative to the nominal disruption radius $R_T$, usually expressed as the
ratio $\beta = R_T/R_p$. Since we are simply adopting $\eta=1$ in Equation~\ref{eqn:disruption},     
$\beta$ is restricted to the range $1 < \beta < \beta_m$ where
\begin{equation}
  \beta_m = { R_T \over R_S } = 5.1 R_{*1} M_{*1}^{-1/3} M_{BH7}^{-2/3}
\end{equation}
is the point where the star passes through the horizon (arguably, we
could also use the last stable orbit).  
The distribution of encounters
in $\beta$ is non-trivial and cannot simply be derived from the 
$ r \propto R_T^{1/4}$ scaling of Equation~\ref{eqn:rate}.  If 
Equation~\ref{eqn:rate} really was directly related to the distribution of pericentric 
radii for disrupting stars, it would imply an unphysical differential
distribution of $dP/dR_p \propto R_p^{-3/4}$ that is dominated by strongly
radial orbits.  The same holds for the $dP/dR_p \propto 1/R_p$
distribution adopted by \cite{Strubbe2009}.

TDE rates are dominated by two limiting regimes (see, e.g., \citealt{Wang2004}).  
For more distant orbits, the 
orbital angular momentum changes relative to the scale needed to pass close 
to the black hole faster than the orbital time scale.  In this ``pinhole'' limit, 
the angular momentum for an encounter is random.  For closer orbits,
the angular momentum changes slowly compared to the orbital time scale.
In this ``diffusion'' limit, an orbit slowly approaches the critical 
angular momentum needed to reach $R_T$.  In the ``pinhole'' limit, 
$dP/dR_p$ is constant once we include the effects of gravitational
focusing, and thus $dP/d\beta \propto \beta^{-2}$ (e.g., \citealt{Luminet1990}).
In the diffusion limit, all stars disrupt very close to $R_T$, so 
$dP/dR_p \propto \delta (R_p - R_T)$ and $dP/d\beta \propto \delta (\beta -1 )$.  
Following \cite{Stone2016} we model this as
\begin{equation}
  {dP \over d\beta} \simeq \left\lbrace
  \begin{array}{ll}
      f_{pin} \beta^{-2} \left(1-\beta_m^{-1}\right) &1 < \beta \leq \beta_m \\
      \left( 1-f_{pin} \right)                       &\beta=1 
  \end{array}
  \right.
  \label{eqn:bdist}
\end{equation}
where a reasonable match to their estimates of the fraction of ``pinhole'' mergers is
\begin{equation}
     f_{pin} \simeq \left( 1 + M_{BH7}^{1/2} \right)^{-1}.
     \label{eqn:fpin}
\end{equation}
Low mass black holes have core structures favoring encounters closer
than $R_T$ in addition to having large $\beta_m$, allowing such 
encounters while remaining outside the black hole. 

The importance of $\beta=R_T/R_p$ is presently under discussion. 
The simplest physical picture of a TDE is that the star reaches 
pericenter intact and then is disrupted with a spread in orbital
binding energy set by the tides across the star, 
\begin{equation}
    \delta \epsilon \simeq { G M_{BH} R_* \over R_T^2 } \beta^n
\end{equation}
with $R_S \leq R_p \leq R_T$ ($1 \leq \beta \leq \beta_m=R_T/R_S$) 
and $n=2$ (following \citealt{Stone2013}).  
However, \cite{Guillochon2013} (also see \citealt{Hayasaki2013})
found that this was not the case in their numerical simulations
and that the spread in energy was essentially independent of 
$\beta$, implying $n=0$.   This was further confirmed by the
semi-analytical study of \cite{Stone2013}.  In essence, 
the star ceases to be bound as it crosses $R_T$ and the debris already
proceeding on independent orbits before it approaches
pericenter.  

The consequence of $n=0$ rather than $n=2$ can be seen in how
the energy spread then determines the characteristic fall back time,
\begin{equation}
    t_{fb} =  0.36 \beta^{-3n/2} M_{BH7}^{1/2} m_{*1}^{-1} R_{*1}^{3/2}~\hbox{years}
    \label{eqn:tfb}.
\end{equation}
For a given fall back time, there is a characteristic peak accretion rate of $\dot{M}_{peak} = M_* c^2/3 t_{fb}$.
Compared to the Eddington rate, this accretion rate is
\begin{equation}
   {\dot{M}_{peak} \over \dot{M}_E } \simeq 4.4 \beta^{3n/2} M_{BH7}^{-3/2} M_{*1}^2 R_{*1}^{-3/2}, 
    \label{eqn:mdot}
\end{equation}
assuming a radiative efficiency of $\eta = 10\%$ ($L = \eta \dot{M} c^2$). 
This is an overestimate of the accretion rate if only (part of) the 
envelope is stripped, as can occur for evolved stars (see \citealt{Macleod2012}).
There is a big difference between the $n=2$ scaling, where
close encounters produce short ($t_{fb} \propto \beta^{-3}$), high peak accretion
rate events ($\dot{M}_{peak} \propto \beta^3$), and $n=0$, where the time scale
and peak accretion rates are independent of $\beta$.  

For surveys, the discovery of a TDE largely depends on $t_{fb}$ and $\dot{M}_{peak}$
-- the rise time must be short enough to trigger a detection, and the peak luminosity
must be high enough to allow detection of events in a large enough volume.
Overall event durations for TDEs are long enough to be unimportant factors for 
discovery.  If $n=2$, the importance of ``pinhole'' encounters is greatly 
enhanced.  First, events with long $t_{fb}$ at $\beta=1$ become short enough
to trigger a transient survey.  Second, and more importantly, if peak 
luminosities determine detectability and scale as 
$L_{peak} \propto \dot{M}_{peak} \propto \beta^3$, then the survey
volume scales as $V \propto L_{peak}^{3/2} \propto \beta^{9/2}$.  Since
$dN/d\beta \propto \beta^{-2}$, the contribution of events at a given $\beta$
scales as $dr/d\beta \propto V dN/d\beta \propto \beta^{1/2}$ and events
with $\beta \simeq \beta_m$ (modulo the Eddington limit) will dominate the
observed rates.\footnote{Note that \cite{Stone2013} use the integral distribution 
$P(>\beta) \propto \beta^{-1}$ in their discussion rather than the 
differential distribution $dP/d\beta \propto \beta^{-2}$.} If, on the
other hand, $n=0$, then the time scales and peak luminosities are 
independent of $\beta$ and the relative detectability of events with
differing $\beta$ must depend on higher order effects than the basic
time and accretion rate scales.  To avoid considering too many cases,
we will follow \cite{Guillochon2013}, \cite{Stone2013} and \cite{Stone2016}
and assume
$n=0$ for our primary discussion in \S3.  We illustrate some 
consequences of $n=2$ in the Appendix.

If we focus on the UV/optical TDEs, the observed optical/UV spectra are broadly consistent
with black bodies (\citealt{Holoien2014}, \citealt{Holoien2016}), 
although the turn over at short wavelengths is not always observed and it is 
clear that some TDEs with thermal optical/UV properties have significant non-thermal
emission components (\citealt{Holoien2016}).  The general assumption for these events is that the observed
emission is reprocessed emission from an underlying accretion disk (e.g.,
\citealt{Loeb1997}, \citealt{Strubbe2009}, \citealt{Guillochon2014},
\citealt{Strubbe2015}, \citealt{Stone2016})
rather than direct emission from an accretion disk (e.g., \citealt{Strubbe2009},
\citealt{Lodato2011}),
although other hypotheses have been advanced (e.g., \citealt{Svirski2015}).  To the extent this
is true and the observed temperature is hot compared to typical survey
bands (e.g. V-band for ASAS-SN), the observed peak luminosity is
\begin{equation}
   \left( \nu L_\nu \right)_{peak} = \epsilon L_{peak} 
       { 15 \over \pi^4 } \left( { h \nu \over k T_{peak} }\right)^3
\end{equation}
where $\epsilon$ is some dimensionless efficiency factor.
This reprocessing model is different from \cite{Lodato2011}, where the optical
emission is the tail of direct emission from an accretion disk. 
To the extent that $T_{peak}$ does not vary wildly as a
systematic function of parameters ($M_{BH}$, $\beta$, etc.), and
$L_{peak}$ is Eddington limited, then there is a very simple model
for the relative survey volumes to be associated with different
events.  If an Eddington limited event
from a $M_{BH}= 10^7 M_\odot$ black hole would be detected in
a local survey out in volume $V_0$, then any other event would be
detected in volume
\begin{equation}
   { V \over V_0 } = \left( { \min\left( \dot{M}_{peak}, \dot{M}_E(M_{BH})\right) \over \dot{M}_E(10^7 M_\odot) } \right)^{3/2}.
   \label{eqn:select}
\end{equation}
This provides a simple approach to reasonably estimating the differential
contributions of events to a survey, while avoiding the very much harder
problem of determining $L_{peak}$/$T_{peak}$ and hence 
the volume $V_0$ in which the fiducial event
would be detectable.  More importantly, $V_0$ can simply be estimated
empirically from the properties of observed transients, as we will
do crudely for the ASAS-SN survey in \S3.3.  
Equation~\ref{eqn:select} is appropriate for
low redshift surveys like ASAS-SN where cosmology and evolution can
be neglected.  Other effects may also suppress the
observed contributions from lower mass black holes. For example, 
\cite{Stone2016} explore a model where only relatively close
encounters ($R_p  < 6 R_S$) circularize the debris rapidly and 
produce a strong flare, which means that only high $\beta$ 
pinhole encounters contribute to the TDE rate as $M_{BH}$
decreases below $\sim 10^7 M_\odot$.

\begin{figure}
\centerline{\includegraphics[width=3.5in]{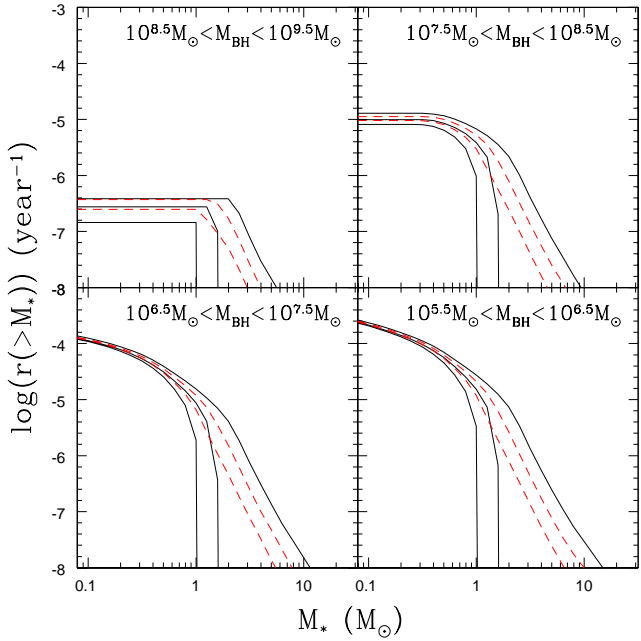}}
\caption{
  Integral TDE rates, $r(>M_*)$, per black hole as a function of stellar mass $M_*$ for black hole mass
  ranges of $10^{8.5}$-$10^{9.5}$ (top left), $10^{7.5}$-$10^{8.5}$ (top right), 
  $10^{6.5}$-$10^{7.5}$ (lower left), $10^{5.5}$-$10^{6.5}$ (lower left) and
  either 1~Gyr burst (black, solid) or constant (red, dashed) star formation models. The
  present ages (from most to fewest higher mass stars) are 1, 3 and 10~Gyr.   
  The two star formation models are identical at 1~Gyr.
  }
\label{fig:demo1}
\end{figure}

\begin{figure}
\centerline{\includegraphics[width=3.5in]{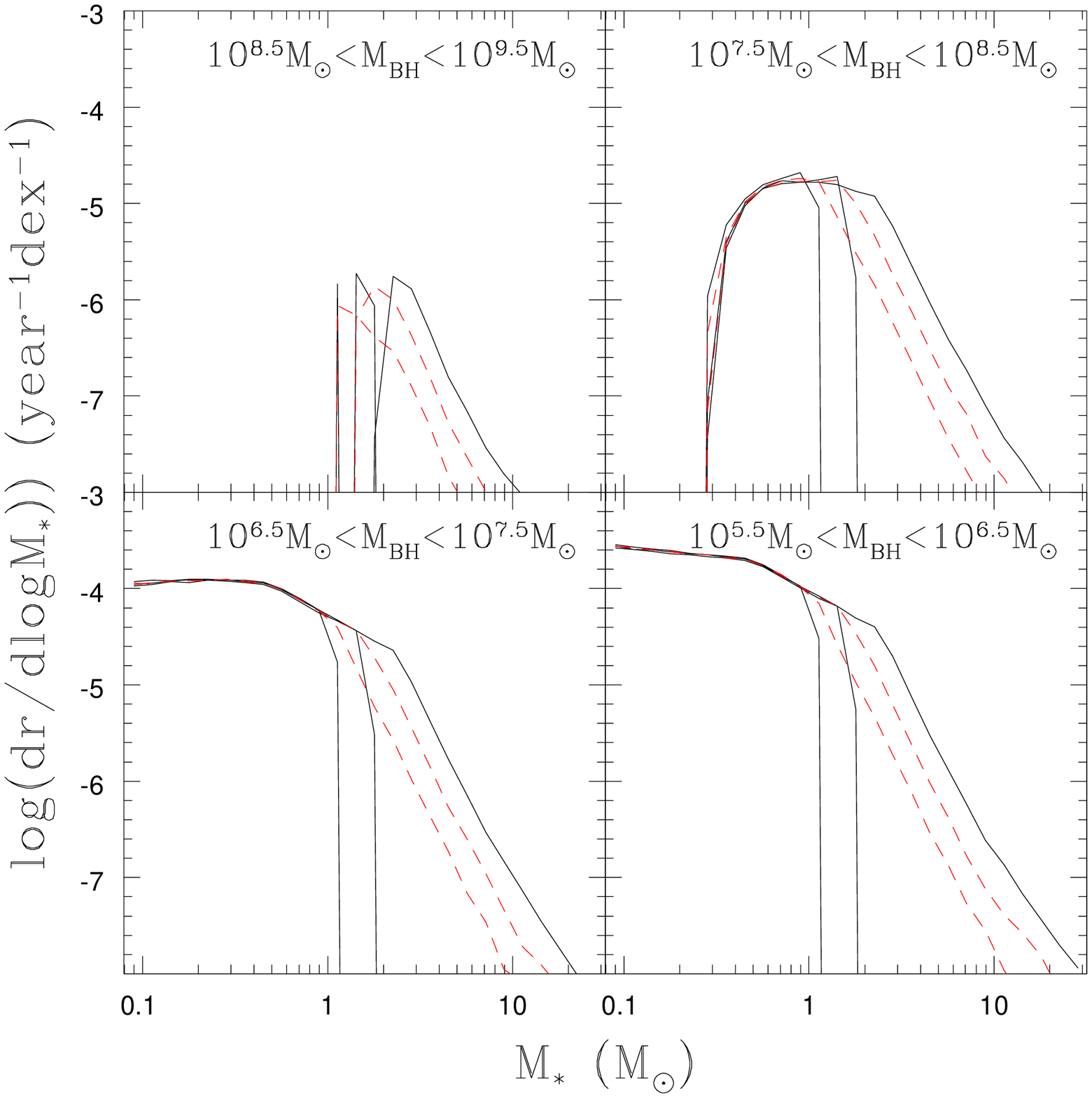}}
\caption{
  Differential TDE rates, $dr/d\log M_*$, per black hole as a function of stellar mass $M_*$ for black hole mass
  ranges of $10^{8.5}$-$10^{9.5}$ (top left), $10^{7.5}$-$10^{8.5}$ (top right), 
  $10^{6.5}$-$10^{7.5}$ (lower left), $10^{5.5}$-$10^{6.5}$ (lower left) and
  either 1~Gyr burst (black, solid) or constant (red, dashed) star formation models. The
  present ages (from most to fewest higher mass stars) are 1, 3 and 10~Gyr.
  The two star formation models are identical at 1~Gyr.
  }
\label{fig:demo1b}
\end{figure}

\begin{figure}
\centerline{\includegraphics[width=3.5in]{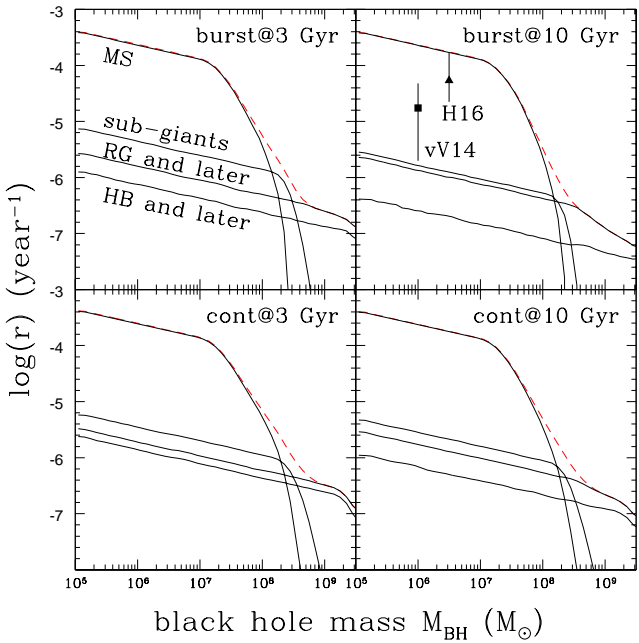}}
\caption{
  Rates per black hole for different stellar evolutionary phases as a function of black hole mass for
  the 1~Gyr burst model at 3~Gyr (top left) or 10~Gyr (top right)
  and the continuous star formation model at 3~Gyr (lower left) or
  10~Gyr (lower right).  The red dashed line shows the total
  rate, and the solid lines show (from top to bottom) the rates
  for MS stars, sub-giants, red giant and later evolutionary states
  and horizontal branch and later evolutionary states. 
  The two points with error bars are the rate estimates 
  by \protect\cite{vanvelzen2014} (vV14) and \protect\cite{Holoien2016} (H16).  
  Their location in black hole mass is in the range of the BH
  mass estimates for observed TDEs.
  }
\label{fig:demo2}
\end{figure}

\begin{figure}
\centerline{\includegraphics[width=3.5in]{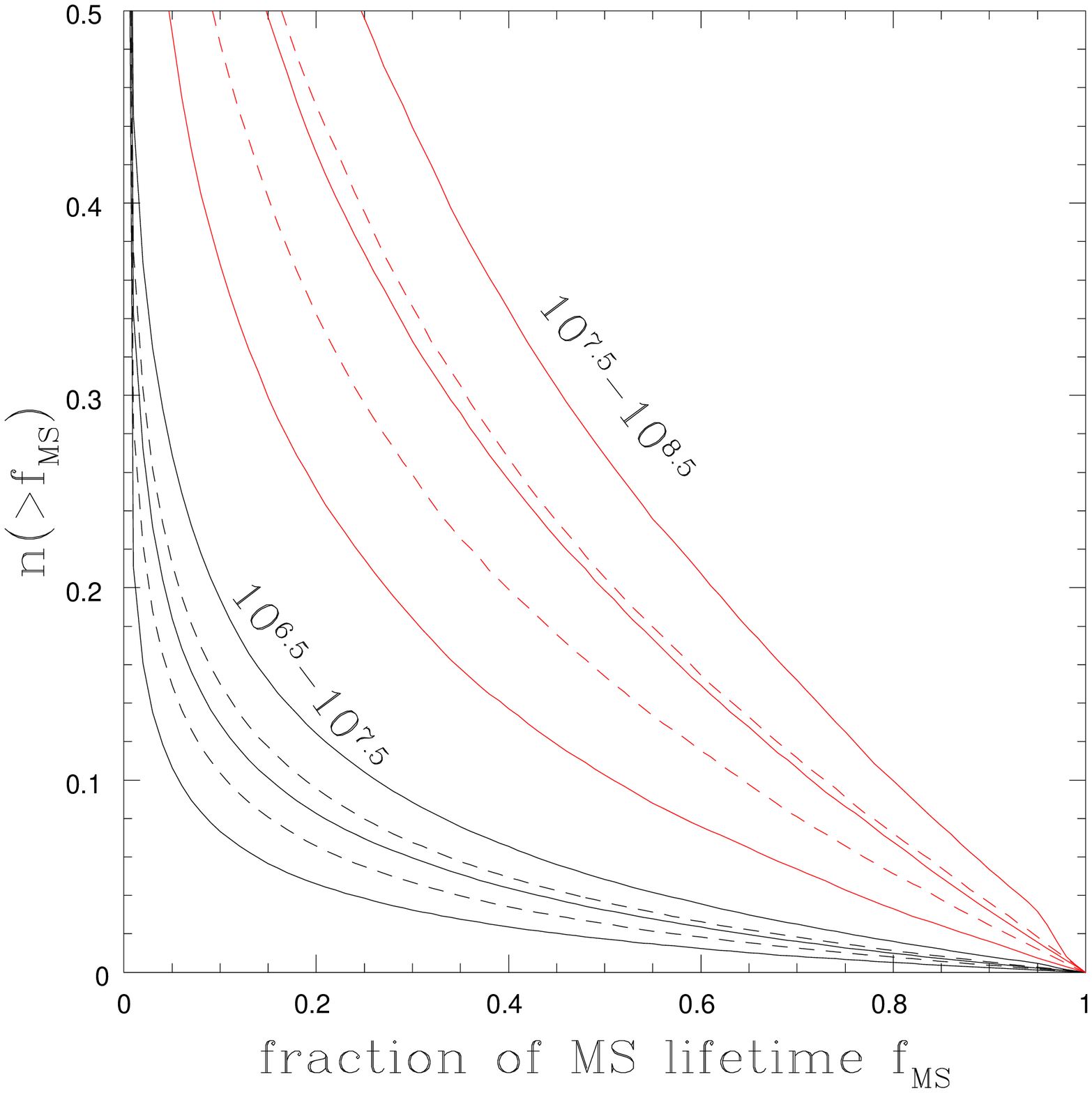}}
\caption{
  Integral distribution $n(>f_{MS})$ of main sequence stars in the fraction
  of elapsed main sequence lifetime $f_{MS}$ at disruption for black
  hole mass ranges of $10^{6.5}$-$10^{7.5}$ (black lower/left) and 
  $10^{7.5}$-$10^{8.5}M_\odot$ (red, upper/right).
  The solid curves are for the 1~Gyr burst model at 1 (lower), 3 (middle)
  and 10~Gyr (top), and the dashed curves are for the constant star
  formation model at 3 (lower) and 10~Gyr (top).  The distributions
  are normalized to $n(>f_{MS}=0)=1$ but the upper halves of the distributions
  are not shown to make the distribution near $f_{MS}=1$ more visible.
  }
\label{fig:demo3}
\end{figure}

\section{TDE Demographics}

In this section we explore various aspects of the demographics of TDEs.  In
\S3.1 we examine the effects of the star formation history on the masses
and evolutionary states of the disrupted stars as a function of 
black hole mass. In \S3.2 we examine the effects changing the 
scaling of the capture rates in Equation~\ref{eqn:rateuse} with
black hole mass.  In \S3.3 we survey the 
properties of local TDEs in stellar mass, evolutionary state, peak accretion
rate ($\dot{M}_{peak}/\dot{M}_E$), fall-back time ($t_{fb}$) and 
pericentric distance ($\beta = R_T/R_p$).  In \S3.4 we examine the
evolution of TDE rates with redshift.

\subsection{Stellar Mass and Evolutionary State}

Figures~\ref{fig:demo1} and \ref{fig:demo1b} show the integral, $r(>M_*)$,
and differential, $dr/d\log M_*$, TDE rates as a function of stellar mass
for black hole mass bins centered at $M_{BH}=10^6$, $10^7$, $10^8$, and 
$10^9 M_\odot$ and the two star
formation histories (a 1~Gyr burst or constant) at ages of $1$, 
$3$ and $10$~Gyr.  At 1~Gyr, the two star formation histories are
identical.  The rate estimates at $M_{BH} \sim 10^6 M_\odot$ of
roughly $10^{-4}$~year$^{-1}$ are consistent with earlier results,
as they must be given their underlying dependence on the \cite{Wang2004}
models.  

For the two lower black hole mass ranges, the rates are dominated by lower
mass, main sequence stars, as was already well known. The
rates increase slightly towards lower black hole masses because of
the weak black hole mass dependence of Equation~\ref{eqn:rateuse}. 
Because the \cite{Kroupa2001} mass function 
is almost logarithmically flat at low masses and it is somewhat easier to disrupt
higher mass, lower mean density main sequence stars, the logarithmic rates
are almost constant for $M_*<0.5 M_\odot$ and then drop slowly
up to $M_* \simeq M_\odot$ due to the break in the slope of the
IMF.  The typical TDE is of an $M_*  \simeq 0.3 M_\odot$
main sequence M-dwarf.  The star formation history completely controls the
rates for higher mass stars, but affects the total rates little
since only 14\% of stars on the IMF have $M_* > M_\odot$.  The
rapid decline in stellar lifetimes with stellar mass leads to 
very sharp breaks in the mass dependent rates for the burst
star formation models at late times.

In the next higher mass range, centered on $M_{BH}=10^8 M_\odot$, we begin to
see the effects of the dropping strength of the tidal gravity at
the event horizon as the black hole mass increases.  The lower
mass M dwarfs can no longer be disrupted, leading to an order
of magnitude drop in the TDE rate. The mass function of the 
disrupted stars is strongly truncated near 
$M_* \simeq 0.3 M_\odot$, so the typical mass of a disrupted
star increases and now depends more on the 
star formation history.  While the existence of the 
sharp cutoff at low masses is generic, its exact location
will be sensitive to the radius used to define the
boundary between disruption and absorption.  \cite{Kesden2012}
explores some of these issues for rotating black holes and
finds that they have only modest effects.  The primary
effect of the star formation history is still to modify
the mass function of the disrupted stars at $M_*\gtorder M_\odot$.
The total rates for the different star formation histories differ by only $\sim 25\%$. 

For the highest mass black holes, the mass function of the disrupted stars
and the overall rates depend strongly on the star formation history because it is increasingly only
evolved stars that can undergo a TDE.  The absolute rates
now vary by a factor of $\sim 3$ between the star formation
histories and the typical disrupted star has the mass of a
star at the MS turn off.  For star formation histories where
the duration of the star formation period is still relevant,
there is a peak with a power-law decline towards higher
masses.  For the older burst populations, the mass function
increasingly looks like a delta function because the range
of stellar ages corresponds to a negligible spread in MS
turn off masses. For the higher mass black holes ($M_{BH} \gtorder 10^8M_\odot$)
it is likely that only the burst models are relevant at lower
redshifts because the hosts will be early-type galaxies with
old stellar populations.

Figure~\ref{fig:demo2} shows the rates as a function of the 
evolutionary state of the star and the black hole mass for
the 3~Gyr and 10~Gyr old stellar population models.  For 
simplicity we simply show the results for all stars, 
MS stars, sub-giants, stars that have evolved past
the base of the RGB, and horizontal branch and later
stars.  \cite{Macleod2012} illustrate 
relative rate distributions over these later evolutionary 
phases for a range of stellar masses.  Figure~\ref{fig:demo2}
also compares the estimates to the two recent rate
estimates by \cite{vanvelzen2014} and
\cite{Holoien2016}.  Neither study differentiates by
black hole mass, but typical black hole mass estimates
for observed optical/UV TDEs are of order $M_{BH}=10^6$ to $10^{7.5} M_\odot$
(see below).  The two rate estimates are in mild conflict,
but they illustrate the continuing tension between observed
and theoretical disruption rates.

For lower black hole masses where all stars disrupt 
($M_{BH} \ltorder 10^7 M_\odot$), the rate steadily
rises for lower masses because of the $M_{BH}^{-1/4}$ 
scaling of the adopted rate (Equation~\ref{eqn:rateuse}).
Disruptions are
overwhelmingly dominated by MS stars, with evolved stars
representing only $\sim 3\%$ of the rate.  The rate
for evolved stars is dominated by sub-giants, then
red giants, and then all later phases.  The average
mass of the disrupted evolved stars will be much 
higher than the main sequence stars, so selection 
effects can significantly modify the observed ratios,
as we will discuss in \S3.3.

Starting around $M_{BH} \gtorder 10^{7.5} M_\odot$, an increasing
fraction of lower mass MS stars are absorbed rather than
disrupted, leading to a very rapid drop in the TDE
rate at higher black hole masses.  At roughly
$M_{BH}=10^8 M_\odot$, the rates for evolved and MS
stars are comparable, with the subgiants still dominating
the rates for evolved stars.  The black hole mass
scale of the rapid drop in the rates will shift in direct
proportion to changes in the criterion for absorption
over disruption.  At slightly higher black hole masses,
the sub-giant contribution also drops rapidly due
to a combination of two factors.  First, subgiants
are not tremendously larger than MS turn off stars,
so for fixed stellar mass it does not take a huge
increase in the black hole mass to absorb a 
sub-giant as compared to a MS turn off star.  Second,
the subgiants are associated with the highest
mass MS stars in the stellar population, because
the lower mass stars have not had time to evolve.
As a result, the sub-giant contribution to the rates
drops rapidly at masses only slightly above the
black hole mass where they are as important as MS
stars.  Finally, as the black hole mass approaches
$M_{BH} \gtorder 10^{8.5} M_\odot$, only the giant stars 
contribute to the rates.  Since such evolved stars
are rare in all star formation histories, the
expected rates are $2$ to $3$ orders of magnitude
lower than for $M_{BH} \ltorder 10^{7.5} M_\odot$ 
where all MS stars will be disrupted.

\cite{Kochanek2015} noted that $M_* \gtorder M_\odot$
stars develop significantly depressed (enhanced) carbon (nitrogen)
average abundances even by fraction $f_{MS} \simeq 0.1$ of their 
MS lifetime due to CNO reactions. Along with their slowly increasing helium
mass fractions, this means that stellar evolution can lead to 
abundance anomalies in TDE debris and potentially their spectra. This is a different
concern from studies of nuclear reactions triggered by
deeply plunging TDE orbits (e.g., \citealt{Luminet1989}). 
Figure~\ref{fig:demo3} shows the integral distribution $n(>f_{MS})$
for the two star formation models and ages of $1$, $3$ and
$10$~Gyr for black hole mass ranges of $10^{6.5}$-$10^{7.5}$
and $10^{7.5}$-$10^{8.5}M_\odot$.  Lower black hole mass
ranges have distributions very slowly shifting to having
fewer evolved stars, while higher mass black hole ranges
quickly shift to being dominated by stars close to the
MS turn off.  For black hole masses $M_{BH} \ltorder 10^7 M_\odot$,
the huge numbers of long lived, low mass MS stars which have
had no time to evolve dominate the TDE rates.  As a result
only some 10-20\% of MS TDEs will have significantly 
anomalous carbon and nitrogen abundances, and only 
$\sim 5\%$ will have $f_{MS}\gtorder 0.5$ where the 
helium enhancement begins to become significant.  To
this can be added the $\sim 3\%$ contribution from 
sub-giants, which are also likely to be fully disrupted
in a TDE.  Later evolutionary states (aside from the rare
AGB stars) are less likely to produce anomalous abundance
signatures because the material inside the hydrogen burning
shell is sequestered in a very high density core that is 
largely decoupled from the envelope.  As the black hole
mass increases, the fraction of TDEs associated with more
evolved stars steadily increases.
 
\begin{figure}
\centerline{\includegraphics[width=3.5in]{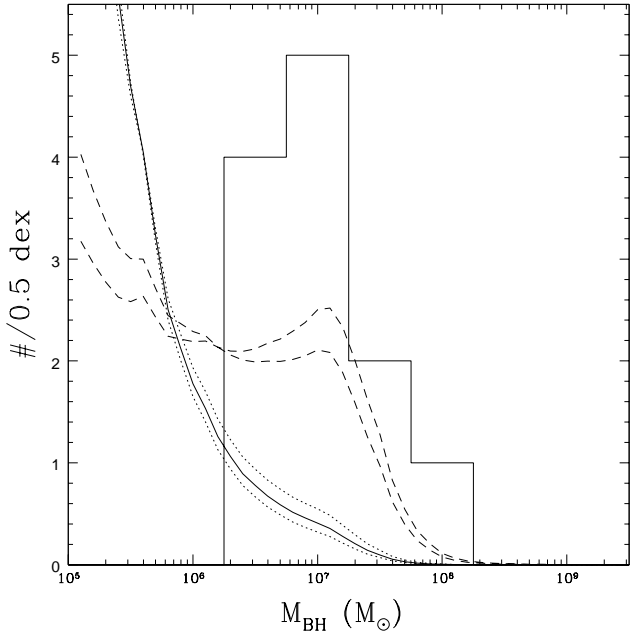}}
\caption{
  Effects of the black hole mass function, $M_{BH}$-$\sigma$ relation
  and the dependence of TDE rates on $M_{BH}$ on the distribution of
  TDEs in black hole mass.  The histogram shows the distribution of black hole mass estimates
  for a sample 12 optical/UV TDEs (see footnote).  The solid black
  curve is our fiducial model based on Equation~\ref{eqn:rateuse}
  and the local \protect\cite{Shankar2009} black hole mass function.
  The dotted lines bracketing it show the effect of varying the
  logarithmic slope of the $M_{BH}$-$\sigma$ relation
  $M_{BH} \propto \sigma^\alpha$ over the range $\alpha = 4.65 \pm 0.50$
  assuming the TDE rate scales as $\sigma^{7/2}$ as in Equation~\ref{eqn:rate}.
  Adopting the much stronger \protect\cite{Brockamp2011} rate scalings
  of $r \propto M_{BH}^{0.35}$ to $M_{BH}^{0.45}$ with black hole mass
  does significantly alter the predictions, as shown by the black
  dashed lines.  These are all for the 1~Gyr burst star formation model
  at an age of 10~Gyr.
  }
\label{fig:msig}
\end{figure}

\subsection{The Local Mass Function and $M_{BH}$-$\sigma$ Relation}

Figure~\ref{fig:msig} shows the expected distribution of local TDEs
as a function of $M_{BH}$
for the 10~Gyr old burst model as a function of black hole mass 
for the \cite{Shankar2009} mass function.  To emphasize how strongly events
from low mass black holes are favored, we have used a linear
scale for the numbers of events per logarithmic mass interval.   
The distributions are normalized for comparison
to a sample of 12 optical/UV TDEs with
black hole mass estimates drawn from the
literature\footnote{We included ASASSN-14ae (\citealt{Holoien2014}),
ASASSN-14li (\citealt{Holoien2016}), TDE1/TDE2 (\citealt{vanvelzen2011}),
PS1~10jh (\citealt{Gezari2012}), PS1~11af (\citealt{Chornock2014}),
PTF09g/PTF09axc/PTF09djl (\citealt{Arcavi2014}), and the GALEX
events from \cite{Gezari2006}, \cite{Gezari2008} and \cite{Gezari2009}.}.
The results for the \cite{Hopkins2007} black hole mass function
are very similar.  

The scaling of the rates with $M_{BH}$ in Equation~\ref{eqn:rateuse}
was something of a compromise.  Many possible changes have little
consequence.  For example, the original rate scaling in Equation~\ref{eqn:rate}
depends on the bulge velocity dispersion as $\sigma^{7/2}$, which
when combined with the $M_{BH}$-$\sigma$ relation 
$M_{BH} \propto \sigma^{4.65}$ in Equation~\ref{eqn:msig}
has a black hole mass dependence of $\sigma^{7/2} \propto M_{BH}^{3/4}$.
Examples of other recent estimates of the exponent 
of the $M_{BH}$-$\sigma$ relation are $5.13$ 
(\citealt{Graham2011}), $4.32$ (\citealt{Schulze2011})
and $5.64$ (\citealt{Mcconnell2013}).
As also shown in Figure~\ref{fig:msig}, varying the 
exponent of the $M_{BH}$-$\sigma$ relation by $\pm 0.5$
has very little effect on the predictions since it changes
the dependence of the rate on black hole mass by a factor
of only about $r \propto M_{BH}^{\pm 0.1}$.  Similarly,
\cite{Stone2016}, based on \cite{Mcconnell2013}, used $r \propto M_{BH}^{-0.404}$, which
leads to changes only slightly larger than the example of
using a steeper $M_{BH}$-$\sigma$ relation.  

The dominance of low mass black holes is largely driven by the 
relatively steep slope of the black hole mass function, 
($n \propto M_{BH}^{-3/2}$ for \citealt{Shankar2009}), rather than the mass dependence
of the rate.  Even a large change in the mass dependence of
the rates has difficulty suppressing the contribution from
low mass black holes.  For example, \cite{Brockamp2011} derive 
rate expressions from a series of N-body experiments that scale as 
$r \propto M_{BH}^{0.446}$ or $M_{BH}^{0.353}$ depending
on their choice of an $M_{BH}$-$\sigma$ relation.  As
shown in Figure~\ref{fig:msig}, this leads to a local
peak near $M_{BH} \sim 10^7 M_\odot$ and a reduced 
but still significant contribution from low mass black holes.  
 
Without
an even stronger black hole mass dependence than 
found by \cite{Brockamp2011}, the divergence of the
volumetric rates for low black hole masses is an
inevitable consequence of the divergent number of
low mass galaxies/halos. 
For the remainder of the paper we simply use our fiducial
model, combining Equation~\ref{eqn:rateuse} with the
\cite{Shankar2009} black hole mass function.  We
show most of the subsequent distributions as a function
of black hole mass, making it relatively easy to 
evaluate the consequences of changing either the
black hole mass function or the black hole mass
dependence of the TDE rates.  We also examine the 
consequences of our simple selection effects model. 

\begin{figure}
\centerline{\includegraphics[width=3.5in]{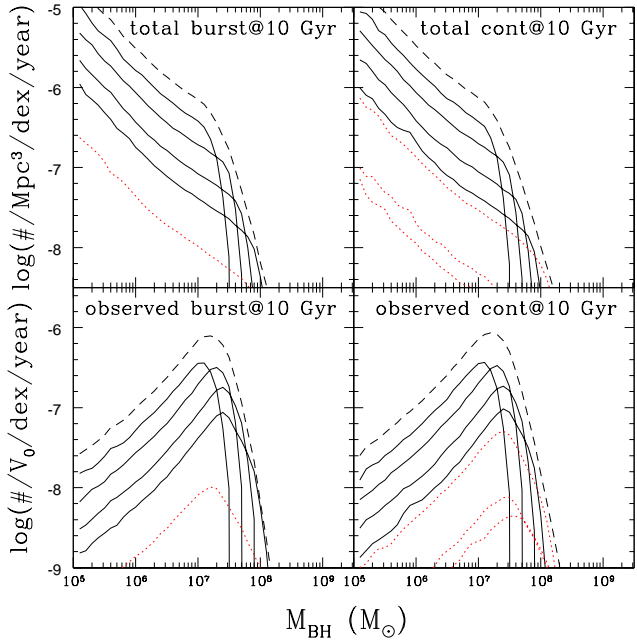}}
\caption{
  Volumetric (top) and observed (bottom) TDE rates as a function of
  stellar mass $M_*$ for the burst (left) and continuous (right)
  star formation models at an age of $10$~Gyr and the local
  \protect\cite{Shankar2009} black hole mass function. In order
  of increasing stellar mass and diminishing TDE rates for low
  black hole masses, the solid curves are for $M_*=0.08$-$0.25$,
  $0.25$-$0.50$, $0.50$-$0.75$ and $0.75$-$1.0M_\odot$, and the
  dotted red curves are for $M_*=1.0$-$1.5$, $1.5$-$2.0$ and $>2.0M_\odot$.
  The dashed curves give the total rate.  In the lower panels
  the rate scales with the fiducial volume $V_0$ (in Mpc$^3$, see text).
  }
\label{fig:smass}
\end{figure}

\begin{figure}
\centerline{\includegraphics[width=3.5in]{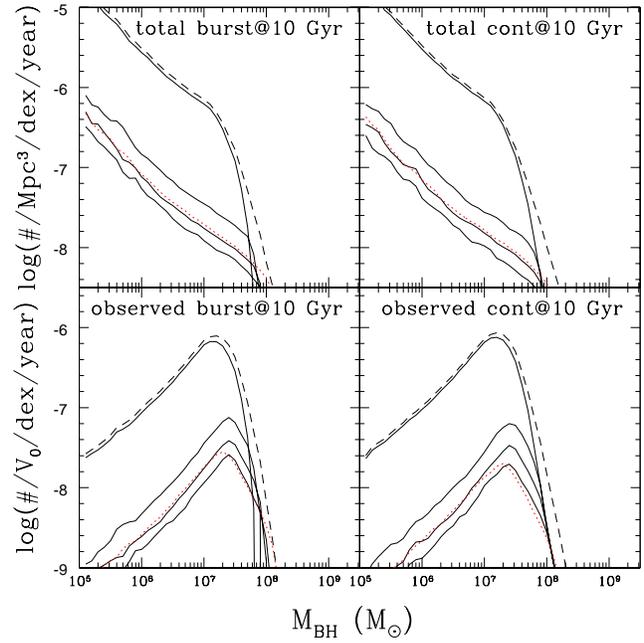}}
\caption{
  Volumetric (top) and observed (bottom) TDE rates as a function of
  stellar evolutionary state for the burst (left) and continuous (right)
  star formation models at an age of $10$~Gyr and the local
  \protect\cite{Shankar2009} black hole mass function. In order
  of increasing stellar age and diminishing TDE rate for low
  black hole masses, the solid curves are for stars with
  main sequence life time fractions of $f_{MS}=0$-$0.25$,
  $0.25$-$0.50$, $0.50$-$0.75$ and $0.75$-$1.00$ and the
  dotted red curves are for post-MS stars.
  The dashed curves give the total rate.  In the lower panels
  the rate scales with the fiducial volume $V_0$ (in Mpc$^3$, see text).
  }
\label{fig:fstar}
\end{figure}

\begin{figure}
\centerline{\includegraphics[width=3.5in]{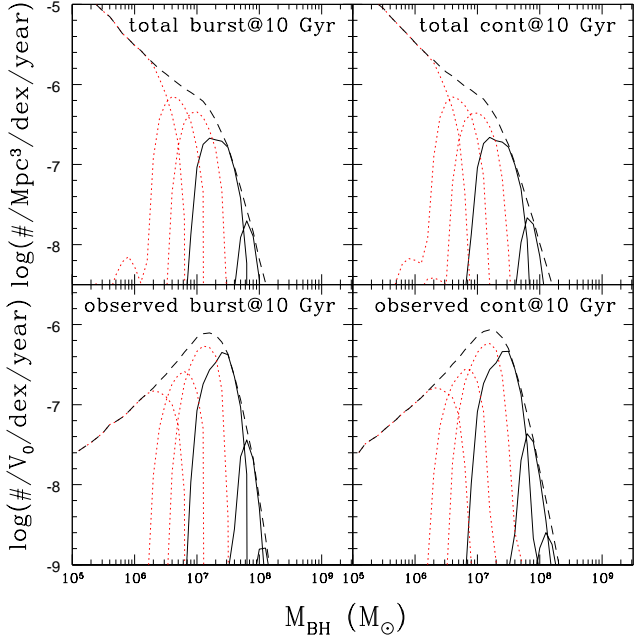}}
\caption{
  Volumetric (top) and observed (bottom) TDE rates as a function of
  the peak accretion rate in Eddington units $\dot{M}_{peak}/\dot{M}_E$
  for the burst (left) and continuous (right)
  star formation models at an age of $10$~Gyr and the local
  \protect\cite{Shankar2009} black hole mass function. In order
  of increasing accretion rate and lower black hole mass, the solid curves are accretion
  rates of $\dot{M}_{peak}/\dot{M}_E<10^{-1.5}$, $10^{-1.5}$-$10^{-1.0}$,
  $10^{-1.0}$-$10^{-0.5}$, and $10^{-0.5}$-$10^{0.0}$ while the
  dotted red curves are for accretion rates of $10^{0.0}$-$10^{0.5}$,
  $10^{0.5}$-$10^{1.0}$ and $>10^{1.0}$.
  The dashed curves show the total rate.  In the lower panels
  the rate scales with the fiducial volume $V_0$ (in Mpc$^3$, see text).
  }
\label{fig:mdot}
\end{figure}

\begin{figure}
\centerline{\includegraphics[width=3.5in]{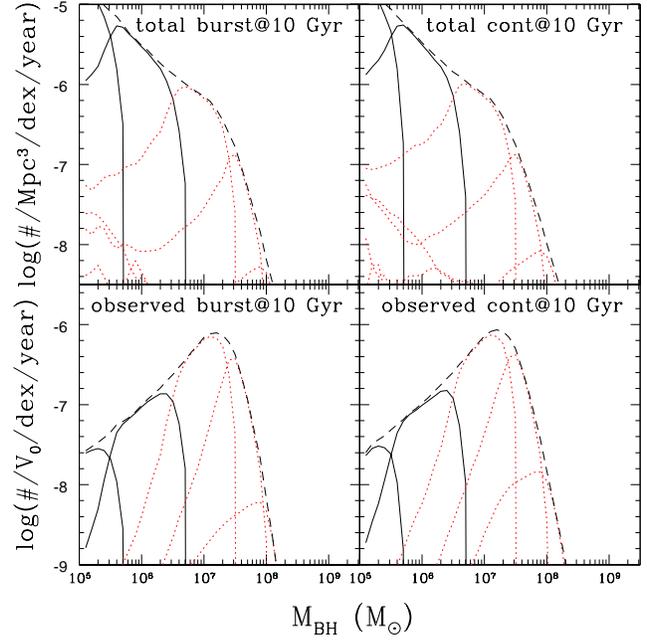}}
\caption{
  Volumetric (top) and observed (bottom) TDE rates as a function of
  the fall back time $t_{fb}$
  for the burst (left) and continuous (right)
  star formation models at an age of $10$~Gyr and the local
  \protect\cite{Shankar2009} black hole mass function. In order
  of increasing accretion rate and (generally) black hole mass, the solid curves are for
  $t_{fb}<10^{-1.5}$~years and $10^{-1.5}$-$10^{-1.0}$,
  while the dotted red curves are for 
  $t_{fb}=10^{-1.0}$-$10^{-0.5}$, $10^{-0.5}$-$10^{0.0}$, $10^{0.0}$-$10^{0.5}$,
  $10^{0.5}$-$10^{1.0}$ and $>10^{1.0}$~years.  Events with
  the time scales of the solid curves are very likely to trigger
  present day transient searches, while the time scales
  corresponding to the dotted red curves are increasingly likely to be ignored.
  Note the contribution of long time scale events for low mass
  black holes due to the disruption of evolved stars.
  The dashed curves show the total rate.  In the lower panels
  the rate scales with the fiducial volume $V_0$ (in Mpc$^3$, see text).
  }
\label{fig:tfb}
\end{figure}

\begin{figure}
\centerline{\includegraphics[width=3.5in]{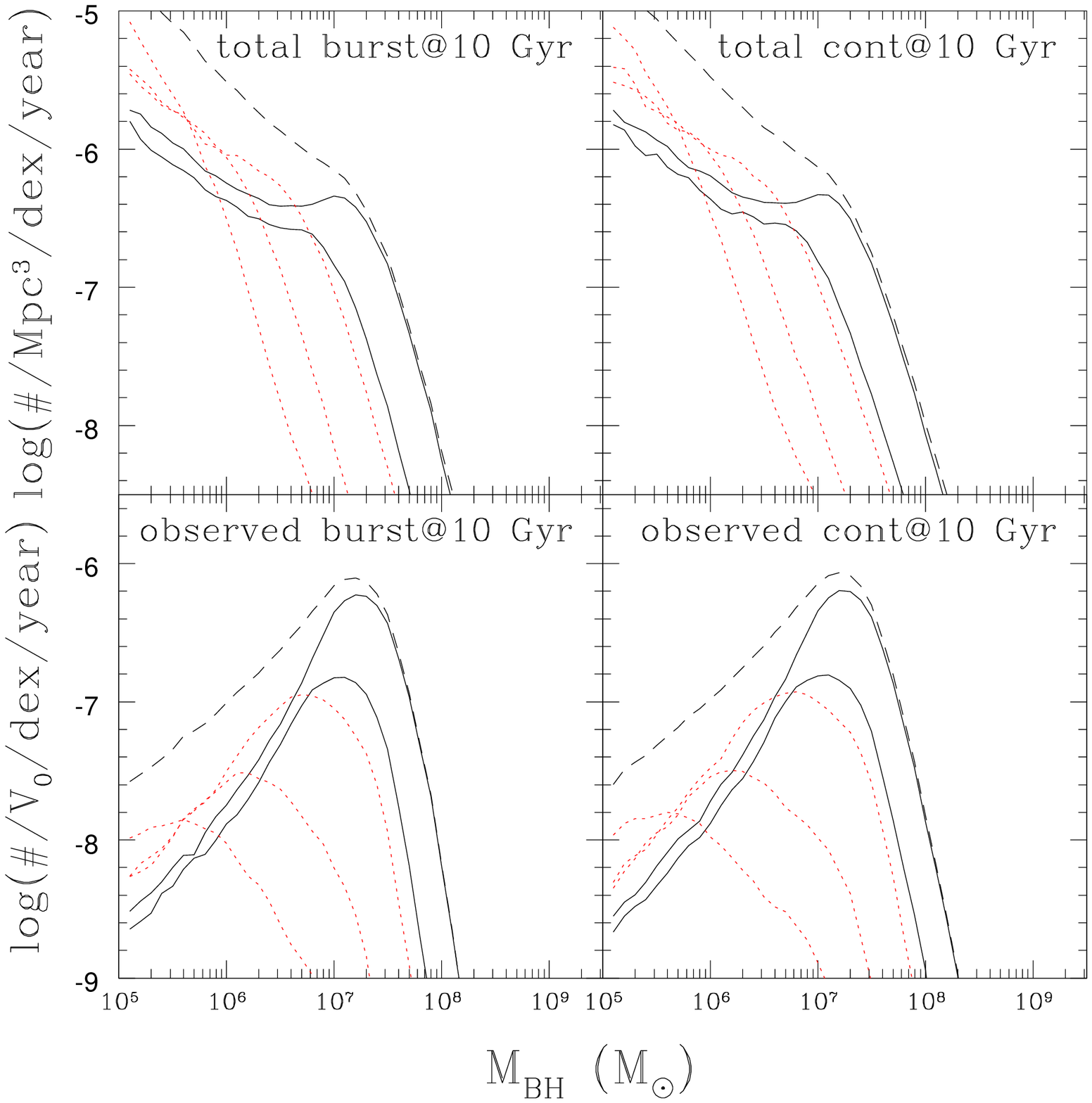}}
\caption{
  Volumetric (top) and observed (bottom) TDE rates as a function of
  the pericentric depth, $\beta = R_T/R_p$, of pinhole encounters
  for the burst (left) and continuous (right)
  star formation models at an age of $10$~Gyr and the local
  \protect\cite{Shankar2009} black hole mass function. In order
  of increasing $\beta$ (smaller pericenters), the solid curves are for
  $\beta=1.0$-$1.5$ and $1.5$-$2.0$,
  while the dotted red curves are for $\beta=2$-$4$, $4$-$8$
  and $>8$.  The dashed curves show the total rate.
  The overall rates must be modified
  by the fraction of pinhole relative to diffusive encounters
  (e.g. Equation~\ref{eqn:fpin}).  In the lower panels
  the rate scales with the fiducial volume $V_0$ (in Mpc$^3$, see text).
  }
\label{fig:beta}
\end{figure}

\subsection{Local Demographics}

Next we explore the distribution of TDE properties as a function
of black hole mass for the local \cite{Shankar2009} mass function
and (as limiting cases) the burst and continuous star formation
models at an age of 10~Gyr.  The results for other star formation
histories can be approximately inferred from the earlier
figures including all 5 star formation histories.  In each
case, we show the ``true'' volumetric rate per logarithmic
black hole mass interval and the ``observed'' rate per unit
fiducial volume $V_0$ (in Mpc$^3$) assuming that $n=0$ so
that high $\beta$ disruptions have the same fall back times 
and peak luminosities as those at $\beta \simeq 1$ (see
the discussion in \S2 and the Appendix).  

As a reminder, the fiducial volume
$V_0$ is nominally the volume in which an Eddington-limited TDE for
a $10^7M_\odot$ black hole would be detected.
We are simply going to give a rough empirical calibration
based on the ASAS-SN TDEs (\citealt{Holoien2014}, \citealt{Holoien2016}),
which are associated with black holes roughly in this mass
range and have peak luminosities that are reasonably close
to the Eddington limit.
For example, ASASSN-14ae (\citealt{Holoien2014}) had
an observed peak of $M_V \simeq -19.5$~mag (the 
true peak was likely slightly higher).  ASAS-SN
can detect such transients out to a comoving 
distance of roughly 200~Mpc and monitors roughly
one-third of the sky after clipping the Galactic
plane and fields that are just rising or setting.
Thus, a reasonable,
empirical estimate for the ASAS-SN survey is
that $V_0 \simeq 10^7$~Mpc$^3$.  This is meant
to provide a rough guide for interpreting rates
rather than as a formal estimate. 

First, in Figures~\ref{fig:smass} and \ref{fig:fstar}, we show the
distributions in stellar mass and evolutionary state. 
In the volumetric rates, we again see that the low mass 
dwarfs completely dominate the rates for 
$M_{BH} \ltorder 10^7M_\odot$, and then the dominant mass
rapidly shifts to higher stellar masses as lower mass stars are
absorbed rather than captured.  For the highest mass black
holes, the volumetric rates are driven by the $M_* \gtorder M_\odot$
stars, but only the continuous star formation models have 
any events from stars significantly more massive than $M_* \simeq M_\odot$.
For our simple selection effects model (Equation~\ref{eqn:select}), the observed contributions from
lower mass black holes are strongly suppressed because the 
Eddington limit on the luminosity restricts the survey 
volume, $V \propto L_{Edd}^{3/2} \propto M_{BH}^{3/2}$. The
contribution at high masses is suppressed by the steeply
falling mass function (Figure~\ref{fig:bhmf}) and because
the  longer fall back time scales and roughly fixed stellar 
masses increasingly limit the accretion rate to be below Eddington. 
As a result, the observed events are strongly peaked near 
$10^7M_\odot$ rather than having the distribution of Figure~\ref{fig:msig}.

Similarly, as long as the rates are dominated by the low mass
dwarfs, most TDEs are from stars that are very young compared 
to their overall MS lifetimes.  The contribution from stars
past the mid-point of their MS lifetimes is nearly $\sim 30$
times lower, For these lower mass black holes, the contributions 
from post-main sequence stars and stars near the end of their 
MS lifetimes are comparable. As the dwarfs are increasingly
absorbed rather than disrupted for higher mass black holes, the distribution of events
in stellar age becomes much more uniform. And then, finally,
the TDEs associated with the highest mass black holes are
increasingly due to evolved stars.  In the observed distributions,
the balance is modestly shifted towards older stars as the
fall back time scales become longer and the numerous low
mass dwarfs can no longer support Eddington-limited accretion.
In both the stellar mass and evolutionary state distributions,
the differences between the burst and continuous star formation
models are relatively subtle, with obvious differences only
for high mass black holes where the TDE rates are dominated by
evolved stars.

Figure~\ref{fig:mdot} and \ref{fig:tfb} show the distributions
of events in peak accretion rate, $\dot{M}_{peak}/\dot{M}_E$, and 
fall-back time, $t_{fb}$, assuming $n=0$ in Equations~\ref{eqn:tfb}
and \ref{eqn:mdot}.  In a 
volume limited sample of TDEs, the spread in the black
hole masses is so large compared to the spread in stellar
masses that there is a tight correlation of $\dot{M}_{peak}/\dot{M}_E$
with $M_{BH}$ even when a stellar mass function is included.
Lower mass black holes have higher accretion rates, and Figure~\ref{fig:mdot}
may underestimate the trend because we simply used $M_*$ for the disruptions
of evolved stars (rather than a partial stripping model, 
e.g., \citealt{Macleod2012}).  If, however, the luminosities
are Eddington limited, then the visibility of the high
$\dot{M}_{peak}/\dot{M}_E$ events associated with lower
mass black holes is greatly reduced, and the sample of
observed TDEs will be fairly tightly clustered around
$\dot{M}_{peak}/\dot{M}_E \sim 1$.  

With the simple selection effects model, the expected observed
TDE rate is roughly $1.5 \times 10^{-6} V_0$/year.  Given our
rough estimate of $V_0 \simeq 10^7$~Mpc$^3$ for ASAS-SN,
this implies a TDE rate in ASAS-SN of order 15/year. In
practice, ASAS-SN is finding roughly one TDE per year
(\citealt{Holoien2014}, \citealt{Holoien2016}).  One
possibility is that our TDE model is overestimating the
rate by an order of magnitude, but the distribution in
fall back times seen in Figure~\ref{fig:tfb} suggests
an alternate explanation.  

The fall back time is largely set by the black hole mass,
and spans a range from 10 days or less at $M_{BH}=10^5 M_\odot$
up to decades at $10^9 M_\odot$.  
If we consider the four lowest time scale
bins in Figure~\ref{fig:tfb}, $t_{fb} < 10^{-1.5}$, $10^{-1.5}$-$10^{-1.0}$, 
$10^{-1.0}$-$10^{-0.5}$ and $10^{-0.5}$-$10^{0.0}$~years
($< 12$~days, 12-37~days, 37-116~days and 116-365~days)
they contain roughly 2\%, 14\%, 61\%, 22\% and 1\% of
the observed events.  To the extent that the fall back
time is a reasonable proxy for event rise times, it is
likely that most transient surveys will increasingly reject
sources with time scales longer than $t_{fb}>10^{-1.0}$ years
that are located at the centers of galaxies because of 
AGN variability and other potential false positives.  If
we required $t_{fb} < 10^{-1.0}$~years as a selection limit, 
we would have only 16\% of the potentially
observable TDEs, leading to a rate of only $2.5$/year
in ASAS-SN that is far more compatible with observed
discovery rate.  This is not a panacea since such a 
cut on the time scales truncates the expected 
black hole mass distribution at a somewhat lower black hole mass
than observed (compare Figures~\ref{fig:msig} and
\ref{fig:tfb}).  Raising the limit on the time scale
to $t_{fb}<10^{-0.75}$~years (65~days) quickly reintroduces
the rate tension, with an expected rate in ASAS-SN of
order 6.7/year.   Nonetheless, the existence of selection
effects related to the fall back time scale seems inevitable.

If such a limit on the fall back time scale is needed to 
reconcile the observed and predicted rates, then it is
probably also necessary for the arguments by \cite{Guillochon2013}
and \cite{Stone2013} that encounters with $\beta >1$ have
similar time scales to those with $\beta=1$ to be correct.
In the Appendix (Figure~\ref{fig:tfb2}), we show the 
consequences of scaling the fall back time with the depth 
of the encounter ($n=2$ in Equation~\ref{eqn:tfb}).  The
distribution of pinhole events across the four lowest time scale
bins is now 26\%, 24\%, 41\% and 8\%, which is a significant
shift towards shorter time scales. While this would be 
diluted by the fraction of encounters that are diffusive
(Equation~\ref{eqn:fpin}), using $n=2$ would make it much
more difficult (impossible?) to use time scale selection
effects to reduce the observed contribution from higher
mass black holes. 

Finally, Figure~\ref{fig:beta} shows the distribution 
in the depth of the encounter, $\beta=R_T/R_p$, as a function
of black hole mass for the pinhole encounters.  This would 
then be diluted by the fraction $1-f_{pin}(M)$ (Equation~\ref{eqn:fpin})
of orbits driven by diffusion to disrupt at $\beta \simeq 1$. For
$M_{BH} \gtorder 10^7 M_\odot$, the vast majority of
pinhole encounters disrupt relatively close to the tidal
limit, with $\beta \ltorder 2$.  Only for relatively low
mass black holes are there significant numbers of deeply
plunging orbits.  The number of plunging encounters in a
volume limited sample is then significantly lower because
the Eddington limit greatly suppresses the survey volume
for the lower mass black holes.  This would not change if
we used $n=2$ instead of $n=0$, so that the plunging encounters
would have significantly shorter fall back times and peak
mass accretion rates (Equations~\ref{eqn:tfb} and ~\ref{eqn:mdot}),
because the luminosities produced by these low black hole masses are
already Eddington-limited. We explore this issue further
in the Appendix. 

\begin{figure}
\centerline{\includegraphics[width=3.5in]{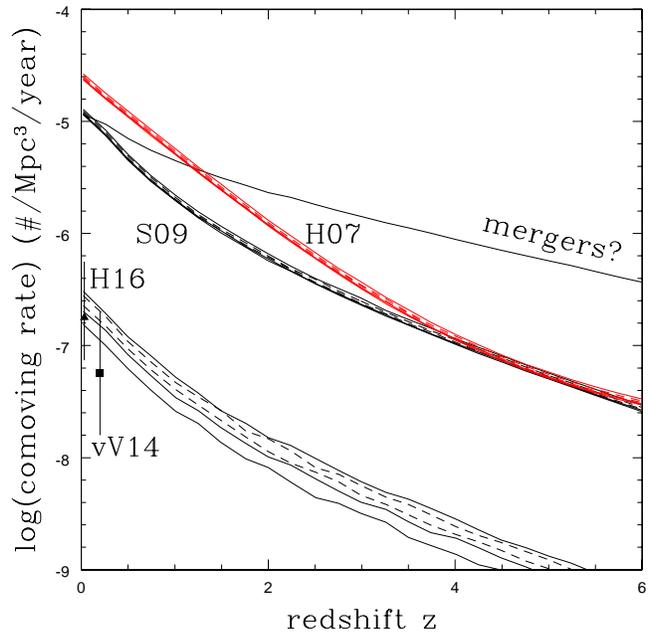}}
\caption{
  The evolution of the comoving TDE rate with redshift for all stars (upper
  curves) and non-MS stars (lower curves).  The standard burst (solid) and
  continuous (dashed) star formation models are shown, where the younger
  models produce more evolved star TDEs.  The total rates are shown for
  both the \protect\cite{Shankar2009} (S09, black) and \protect\cite{Hopkins2007}
  (H07 red) black hole mass functions.  The ``mergers'' curve shows the result
  for a TDE rate driven by mergers with a merger rate scaled by the inverse
  of the age of the universe, $t_H(z)^{-1}$, and normalized to equal the
  contribution of isolated black holes at $z=0$.
  The square point and error bar show
  the volumetric rate estimate by \protect\cite{vanvelzen2014} located
  at the mean redshift of the two candidate TDEs in their sample. The
  triangle and error bar shows the \protect\cite{Holoien2016} rate implied by simply
  maintaining the rate ratio relative to \protect\cite{vanvelzen2014}.
  The stellar populations are not
  evolved with redshift for simplicity since they have so little effect
  on the overall rates compared to the evolution of the black hole
  mass function.
  }
\label{fig:evol2}
\end{figure}

\begin{figure}
\centerline{\includegraphics[width=3.5in]{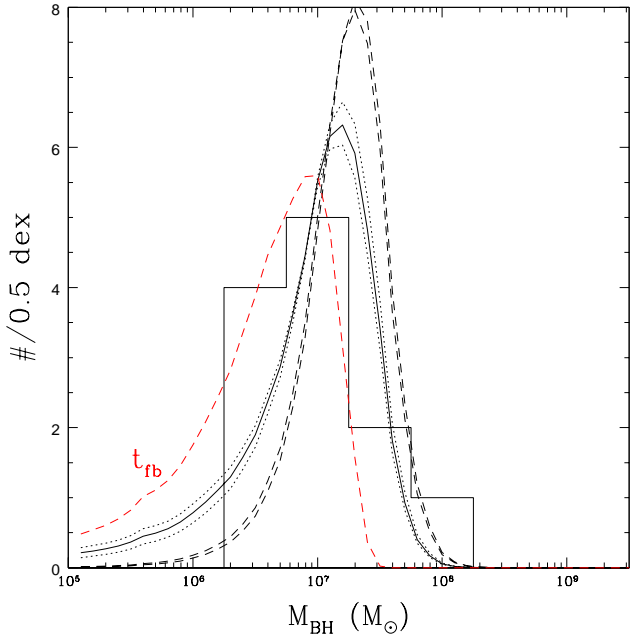}}
\caption{
  Expected distribution of observed TDEs in black hole mass given our simple
  model for selection effects.  As in Figure~\protect\ref{fig:msig}, the
  histogram shows the distribution of black hole mass estimates for a sample
  of 12 optical/UV TDEs.  The solid black
  curve is our fiducial model based on Equation~\ref{eqn:rateuse}
  and the local \protect\cite{Shankar2009} black hole mass function.
  The observed distribution can be well-explained by selection effects
  despite the divergence of the total volumetric rate at low
  black hole mass.
  The dotted lines bracketing it show the effect of varying the
  logarithmic slope of the $M_{BH}$-$\sigma$ relation
  $M_{BH} \propto \sigma^\alpha$ over the range $\alpha = 4.65 \pm 0.50$
  assuming the rate scales as $\sigma^{7/2}$ as in Equation~\ref{eqn:rate}.
  Adopting the much stronger \protect\cite{Brockamp2011} rate scalings
  of $r \propto M_{BH}^{0.35}$ to $M_{BH}^{0.45}$ with black hole mass
  does significantly alter the predictions, as shown by the black
  dashed lines.  The red dashed curve labeled $t_{fb}$ shows an example
  of the effect of excluding events with long fall back times from the
  standard model (see text).
  These are all for the burst star formation model
  at an age of 10~Gyr.
  }
\label{fig:select}
\end{figure}

\begin{figure}
\centerline{\includegraphics[width=3.5in]{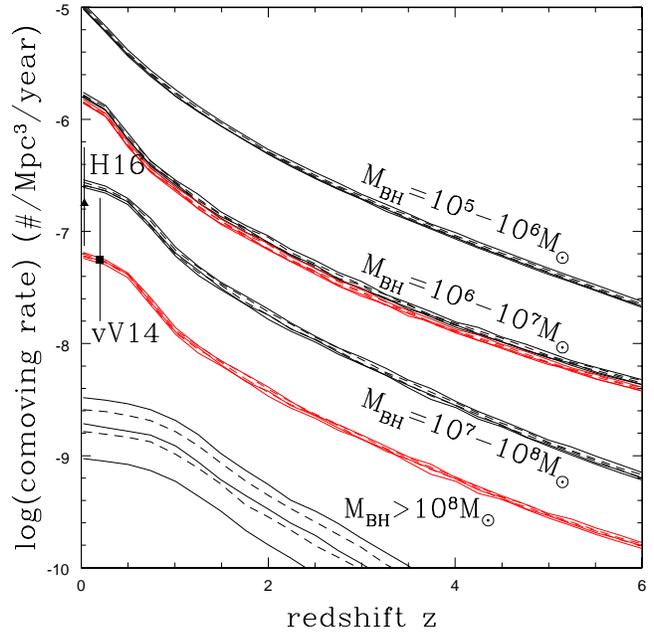}}
\caption{
  The evolution of the comoving TDE rate with redshift in black
  hole mass bins of $M_{BH}=10^5$-$10^6M_\odot$ (top), $M_{BH}=10^6$-$10^7M_\odot$,
   $M_{BH}=10^7$-$10^8M_\odot$, and $M_{BH}>10^8M_\odot$ (bottom) for
  the \protect\cite{Shankar2009} black hole mass function.  The standard burst (solid) and
  continuous (dashed) star formation models are shown, where the younger
  models produce more evolved star TDEs.  The red curves show the effect
  on the volumetric rate of excluding events with long fall back times
  from the standard model for the $M_{BH}=10^6$-$10^7M_\odot$,
  and  $M_{BH}=10^7$-$10^8M_\odot$ mass bins (see text).    
  The point and error bar show
  the volumetric rate estimate by \protect\cite{vanvelzen2014} located
  at the mean redshift of the two candidate TDEs in their sample.
  }
\label{fig:evol3}
\end{figure}

\subsection{The Evolution of TDE Rates With Redshift}

The final issue we explore is the evolution of TDE rates with
redshift.  This was incorporated in the \cite{Strubbe2009} models
using the \cite{Hopkins2007} model for the black hole mass function.
\cite{Strubbe2009} noted that the rates are predicted to 
decline with increasing redshift but did not explore the issue
in any detail.  For simplicity, we will not attempt to simultaneously evolve
the stellar populations with the black hole mass functions.
We know from the earlier sections that the effects of star
formation histories on the rates are small compared to the
rapid changes in the black hole mass function with cosmic epoch 
seen in Figure~\ref{fig:bhmf}.
We will show results for the various star formation histories, and
the consequences of adding any evolution can be understood by 
simply interpolating between curves.  

Figure~\ref{fig:evol2} shows evolution of the rates 
integrated over the $10^5M_\odot < M_{BH} < 10^{9.6}M_\odot$
mass range spanned by the \cite{Shankar2009} models. The
absolute rates are limited by the cutoff in the mass functions
at $M_{BH}=10^5M_\odot$ because of the steep low mass slope
of the mass function (see Figure~\ref{fig:bhmf}).
With the estimated rates dropping by a factor of 5 even by $z=1$, TDEs
are clearly creatures of the local universe.  The \cite{Hopkins2007}
black hole mass function, shown only for the total rates in
Figure~\ref{fig:evol2}, predicts a moderately higher local
rate and a similar decline with redshift.  

For comparison, we also show the volumetric rate estimate by \cite{vanvelzen2014}
located at the mean redshift ($\langle z \rangle =0.20$) of the two candidate TDEs
($z=0.136$ and $z=0.256$, \citealt{vanvelzen2011}).  Like 
their estimated rate per black hole, their observational 
estimate is low compared to our theoretical model (and previous models). \cite{Holoien2016}
did not estimate a volumetric rate, but the mean redshift
of the ASAS-SN TDEs is significantly lower ($\langle z \rangle =0.032$).    
We can approximately convert their rate to a volumetric rate by simply
assuming that the ratio of the volumetric rates between \cite{Holoien2016}
and \cite{vanvelzen2014} is the same as the ratio of the rates per
galaxy shown in Figure~\ref{fig:demo2}.  This rescaling is roughly
consistent with ASAS-SN finding roughly 1 TDE/year in a survey 
volume of roughly $V_0=10^7$~Mpc$^3$ (the estimate we used in \S3.3).
The difference between these estimates (whatever the flaws) is 
consistent with the rapid evolution of TDE rates even if the 
absolute scale differs from the models by a large factor.

That the observed TDEs tend to have $M_{BH} \gtorder 10^6 M_\odot$
(see Figure~\ref{fig:msig}) suggests that the contributions
from the lower mass systems that dominate the theoretical rates
either do not exist because the mass functions are wrong or because
these events are being missed due to selection effects.  Figure~\ref{fig:select}
reprises the $M_{BH}$ distribution in Figure~\ref{fig:msig}
after including our simple model for selection effects.
Despite the divergence of the volumetric rates towards
lower masses, the concentration of the
observed TDEs near $M_{BH} \simeq 10^7 M_\odot$ is
well-reproduced.  The divergence of the mass function
($dn/dM_{BH} \sim M_{BH}^{-3/2}$) is balanced by the
reduction in the survey volume associated with the 
Eddington limit ($V \propto M_{BH}^{+3/2}$) leaving
a distribution $dr/d\log M_{BH} \propto M_{BH}^{3/4}$
for low black hole masses.  Given the uncertainties in
black hole masses, the modest offset between the model
and observed peaks is probably not significant.

Figure~\ref{fig:select} also shows the consequences of adding
a selection limit on $t_{fb}$ to the standard model based on
the discussion in \S3.3.  We use a detection probability of
unity for $t_{fb} < 10^{-1.0}$~years, zero for 
$t_{fb} > 10^{-0.5}$~years and a linear transition in
$\log t_{fb}$ in between.  This model better matches
the peak of the observed TDEs and is still statistically
compatible with observing no lower mass black holes in
the observed sample (although not by much).  It is not 
compatible with the presence
of the higher mass black holes in the observed sample.  
Given that black hole mass estimates are logarithmically
uncertain at the level of $\sim 0.5$~dex, one solution would
be to argue that the high mass tail is simply due to 
scatter in the mass estimates.  Alternatively, we 
could change the $t_{fb}$ selection model to cut off
at somewhat longer time scales or only drop to a 
finite floor instead of zero.  Given the available data,
it presently seems sufficient to simply present an example
of the consequences of including a time scale selection effect.

An alternative view of the role of the black hole mass function
is shown in Figure~\ref{fig:evol3}. Here we show the
rate contributed by four different bins of black hole
mass.  The radical disagreement between the theoretical
and \cite{vanvelzen2014} or the (scaled) \cite{Holoien2016} 
rates in Figure~\ref{fig:evol2} is dominated by the contribution
from the $M_{BH}=10^5$-$10^6M_\odot$ bin.  The differences
between the observed rates and those for the $M_{BH}=10^6$-$10^7M_\odot$ 
black hole mass bin, where the
bulk of the observed TDEs appear to lie (see Figure~\ref{fig:msig}
and \ref{fig:select}), are significantly smaller. 
There is again the huge drop off in rates above
$M_{BH} \gtorder 10^8 M_\odot$.  The black hole mass 
dependence of the TDE rates recapitulates the phenomenon
of ``downsizing'', where the contribution from higher mass
black holes stops increasing rapidly towards lower redshifts 
earlier than for lower mass black holes.

Figure~\ref{fig:evol3} also shows the effects of adding our
simple selection limit on $t_{fb}$ for the $M_{BH}=10^6$-$10^7M_\odot$
and $M_{BH}=10^7$-$10^8M_\odot$ mass bins (it has no
effect on the lowest mass bin and eliminates all the
very high mass events, so these are not shown).  While
adding this limit greatly suppresses the rates for the 
higher mass black holes, it does not have a significant effect on the
rates for the $M_{BH}=10^6$-$10^7M_\odot$ mass range.
Shifting the $t_{fb}$ limits to sufficiently short time
scales would greatly reduce the rates for this lower mass
bin, but would greatly exacerbate the tension between the
observed and predicted black hole mass distributions 
discussed earlier.

Finally, \cite{Arcavi2014} (also see \citealt{French2016}) noted that a surprising number of TDEs
seem to be in post-starburst galaxies (a.k.a. E$+$A or
K$+$A galaxies), suggesting that
TDE rates may be significantly boosted after a merger.
Mergers could do so through two mechanisms.  The first
possible mechanism is that the orbital diffusion time
scales are significantly shorter for an extended period
of time after a merger.  Presumably, mergers do rapidly
refill loss cones, essentially by shifting from orbital
evolution driven by relaxation to evolution driven by
violent relaxation.  
A second, more likely mechanism, is that the post-starburst
phase is also the period when the black holes associated
with the mergers are themselves merging.  For example,
\cite{Li2015} find that TDE rates can be enhanced by 
two orders of magnitude in the phases before the system
becomes a compact binary, although this phase does not 
last 1~Gyr. 

For present purposes, we need no detailed knowledge of
the mechanism to explore the possible consequences of
mergers for the evolution of the rates of TDEs.  
Besides the ``reference'' model we have used here, \cite{Shankar2009}
considered models with mergers.  For merger rates which were relatively
high, there was little effect until $z<1$ where growth by accretion
begins to slow.  Then at lower redshifts, the mergers drove the black
hole mass function to be more dominated by high mass black holes, a
difference which would tend to reduce the TDE rate.  In the \cite{Shankar2009}
models, the rate of mergers is independent of black hole mass
and proportional to the inverse of the age of the universe, $t_H(z)^{-1}$. 
This means that the TDE rate contributed by mergers approximately evolves
as $r(z)t_H(0)/t_H(z)$, where we approximate $r(z)$ using the
fiducial, no merger model of \cite{Shankar2009} since the 
model with mergers was
not tabulated.  Figure~\ref{fig:evol2} shows how such a contribution
to the TDE rate would evolve with redshift, normalized to match
the reference model at $z=0$.  The TDE rate still drops with
redshift, but more slowly because the decrease in the absolute
numbers of black holes (Figure~\ref{fig:bhmf}) is partly 
balanced by the increasing rate of mergers.

\section{Discussion}

As already known, TDE rates should be dominated by the
disruption of low mass ($M_* \ltorder M_\odot$), main 
sequence stars by black holes of modest mass, 
$M_{BH} \ltorder 10^{7.5}M_\odot$.  In detail, the
mass function of disrupted lower mass stars is fairly
flat because the slope of the stellar mass function
is partly balanced by the lower densities of higher mass
stars.  Some $\sim 10\%$ of disrupted stars are 
high enough mass for the CNO cycle to be relevant
and have lived long enough ($f_{MS} \gtorder 10\%$ of their
main sequence lifetime) to show depressed carbon
and enhanced nitrogen abundances, as discussed 
in \cite{Kochanek2015}.  Smaller numbers
of older ($f_{MS} \gtorder 50\%$) main sequence
stars and subgiants will show significant increases
in their average helium abundance.
For higher mass black holes, where the lowest mass
stars are increasingly absorbed rather than disrupted,
the typical mass and age of the disrupted star increases
rapidly.  Above roughly $M_{BH} \gtorder 10^{8.5}M_\odot$,
only evolved stars are disrupted.  
The detailed star formation history is not very 
important except for the the rates associated with
$M_* \gtorder M_\odot$ or evolved stars and high
mass black holes. 
  
The black hole mass function is the crucial factor in
determining TDE rates because simple models of black
hole growth and evolution (e.g. \citealt{Hopkins2007}, 
\citealt{Shankar2009}) predict diverging numbers of 
lower mass black holes just as there are diverging
numbers of low mass halos or galaxies. \cite{Stone2016}
explored this in a more empirical model based on simply
adding black holes to a galaxy luminosity function 
and then simply truncating the mass function below
some limiting black hole mass.  For our black hole 
mass functions, which extend down to $10^5 M_\odot$,
the disagreement between the model and observed rates
(e.g., \citealt{vanvelzen2014}, \citealt{Holoien2016})
is large (factor of $\sim 30$).  One possibility,
explored by \cite{Stone2016}, is that the black hole
mass function is simply truncated below 
$M_{BH} \ltorder 10^6 M_\odot$.  

A second possibility is that selection effects are strongly affecting
the observed rates.  Even with a diverging mass function,
TDEs due to low mass black holes are almost certainly associated
with lower peak luminosities and hence smaller survey volumes.
With small numbers of known TDEs, this can quickly lead to 
discovering no examples of the most common events and hence
gross underestimates of the true volumetric rates.  In particular,
the observed distribution of optical/UV TDE black hole masses is cut off
at low black hole masses in agreement with our simple model
for survey selection effects.  The model also predicts that 
the observed sample should be dominated by somewhat higher mass
black holes than observed, and this is may be explained by a
selection bias against surveys examining slowly rising transients at
the centers of galaxies.  Combining these selection effects 
appears to largely, but not perfectly, reconcile
the observed and theoretical rates.  Searching for TDEs  
with longer time scales in somewhat more massive galaxies 
is likely to be profitable.

The typical accretion rates associated with TDEs are
of order $10^{-5}$ to $10^{-4} M_\odot$~year$^{-1}$,
very similar to the estimate by \cite{Magorrian1999}.
Even without any dramatic evolution in the rates, this means that
the net contribution of TDEs to the mean mass of a black
holes is $10^5$ to $10^6M_\odot$.  Only for low mass black holes
is this rate high enough to significantly contribute to
their growth.  However, the contribution to the growth of the
lower mass black holes comes from disrupted stars rather than
absorbed stars, and it increasingly
appears that most (all?) TDEs accrete very little of the
available mass (see, e.g., the discussion in \cite{Metzger2015}).
If 10\% or less of the mass is actually
accreted, then the TDE process quickly becomes unimportant
for the growth of even the lower mass black holes.  These estimates also
assume that the total mass of the star is disrupted --
for evolved stars with large core/envelope density differences,
it is likely that only the envelope is lost (or portions of
the envelope, see, e.g., \citealt{Macleod2012}).

In the \cite{Hopkins2007} and \cite{Shankar2009} models
for the evolution of the black hole mass function, TDE
rates are predicted to drop very rapidly with redshift,
falling by a factor of $\sim 5$ by a redshift of unity.
The effect is largest for the lowest mass black holes,
in a TDE version of ``downsizing''.  The evolution is
so rapid, that it could explain a significant fraction 
of the factor of $\sim 3$ nominal rate difference between
\cite{vanvelzen2014} at $z \simeq 0.2$ and 
\cite{Holoien2016} at $z \simeq 0.03$. Although it
is also true that the two rate estimates are sufficiently
uncertain to be mutually consistent without the rapid
redshift evolution.  Without a well-established model
for the behavior of TDEs as a function of mass that 
can be used to model selection effects, it may prove
difficult to use TDEs to probe the black hole mass function
as proposed by \cite{Stone2016}.
Since the evolution with redshift at fixed black hole mass
is predicted to be so rapid and TDE properties at fixed black
hole mass are unlikely to evolve rapidly, it should be 
considerably easier to test models for the evolution of
the black hole mass function with redshift using TDEs.
For example, the slowly evolving
theoretical black hole mass function of \cite{Sijacki2015}
used by \cite{Metzger2015b} to predict the rates of jetted
TDEs should be easily distinguishable from the more 
empirical and rapidly evolving models of \cite{Hopkins2007} and 
\cite{Shankar2009}.

While the volumetric rates, which depend on the black hole
mass function, show the largest mismatch to observational
estimates, there may still be significant discrepancies 
between predicted and observed rates per galaxy.  Any
problems have to lie in the rate estimates more than 
uncertainties in other aspects of the estimates such as
the choice of $M_{BH}$-$\sigma$ relations. There is 
less of an issue if the higher rates found by \cite{Holoien2016},
(as compared to \cite{vanvelzen2014}, for example) are
correct.  There are, for example, fundamental differences
between the rate estimates by \cite{Wang2004} or their
updates in \cite{Stone2016} and the estimates of 
\cite{Brockamp2011}.  In the former estimates, the TDE
rate declines with black hole mass ($\propto M_{BH}^{-0.16}$ 
to $\propto M_{BH}^{-0.40}$), while in the latter estimates,
the rate increases with black hole mass ($\propto M_{BH}^{0.35}$
to $M_{BH}^{0.45}$).  Since the range of black hole masses spans
4-5~dex or more, such large differences in the mass scaling
will have an enormous impact on the mass-dependent rates. 
In particular, the rapid rise in the rate with black hole mass
found by \cite{Brockamp2011} greatly suppresses the rate 
contribution of even a divergent population of low mass
black holes.  

More broadly, the semi-empirical rate estimates based on the observed 
structures of galaxies (e.g. \citealt{Magorrian1999}, \citealt{Wang2004}, 
\citealt{Stone2016}), are essentially all for galaxies 
containing black holes above $M_{BH} \gtorder 10^6 M_\odot$.
Part of the problem is the difficulty in identifying systems
with lower mass black holes (see, e.g., \citealt{Greene2007}).
Independent of this problem, present extrapolations of TDE rates 
to lower black hole masses assume an extrapolation of the 
dynamical structure of the host galaxies to lower mass hosts. 
The degree to which this is valid
is an open question, as illustrated by the ongoing debates
about bulges, pseudo-bulges, and black holes 
(see, e.g., \citealt{Kormendy2013},
\citealt{Kormendy2016}, for reviews of some of the issues). 
Detailed models of loss cone dynamics for these lower mass
galaxies are required.

If post-starburst galaxies produce a significant fraction of
TDEs (\citealt{Arcavi2014}, \citealt{French2016}), then the
standard rate estimates are unlikely to be applicable to these
systems.  Many of the TDE properties explored in this paper
may be little affected.  At least in the survey of such galaxies
by \cite{Quintero2004}, the luminosity function of post-starburst
galaxies is almost identical in shape to the luminosity function
of all galaxies -- roughly 1\% of galaxies at all luminosities
show the strong Balmer absorption features characterizing this
class of galaxy.  This suggests that the black hole mass function
of such galaxies is similar to that of all galaxies, and that
just the TDE rate is being accelerated.  However, secondary issues      
(e.g. mass-dependent black hole merger rates if the rate increase 
is due to binary black holes) could then modulate the mass 
dependence of the rates.  Of course, the fundamental tension
between the predicted and observed rates is only exacerbated
if many of the observed TDEs are due to a different mechanism
than the standard loss cone mechanism. 

On the observational side, the key issue is the extent to which
the observed properties of TDEs and their hosts are dominated by
selection effects.
Understanding this requires transient surveys to 
be sensitive to TDEs with (rise) time scales of many months
to years in order to probe higher black hole masses.  If the
suggestion by \cite{Kochanek2015} that nitrogen rich quasars
may be related to TDEs is correct, than surveys for spectral
evolution in (particularly the nitrogen rich) quasars may be
an alternate probe for higher mass TDEs.  Transient surveys
are generally sensitive to the much shorter time scales 
expected for lower mass systems, and may just be limited 
by reduced survey volumes because lower mass black holes 
have lower peak luminosity transients.  Alternatively, if TDE
phenomenology (e.g., X-ray, UV or optically dominated) 
depends strongly on black hole mass, then it will be necessary
to unify the results of all the different search strategies
used to date.

\section*{Acknowledgments}
We thank B. Metzger, N. Stone, T. Thompson and D. Weinberg for discussions.
CSK is supported by NSF grants AST-1515876 and AST-1515927.

\appendix

\section{Pinhole Encounters When $n=2$}

\begin{figure}
\centerline{\includegraphics[width=3.5in]{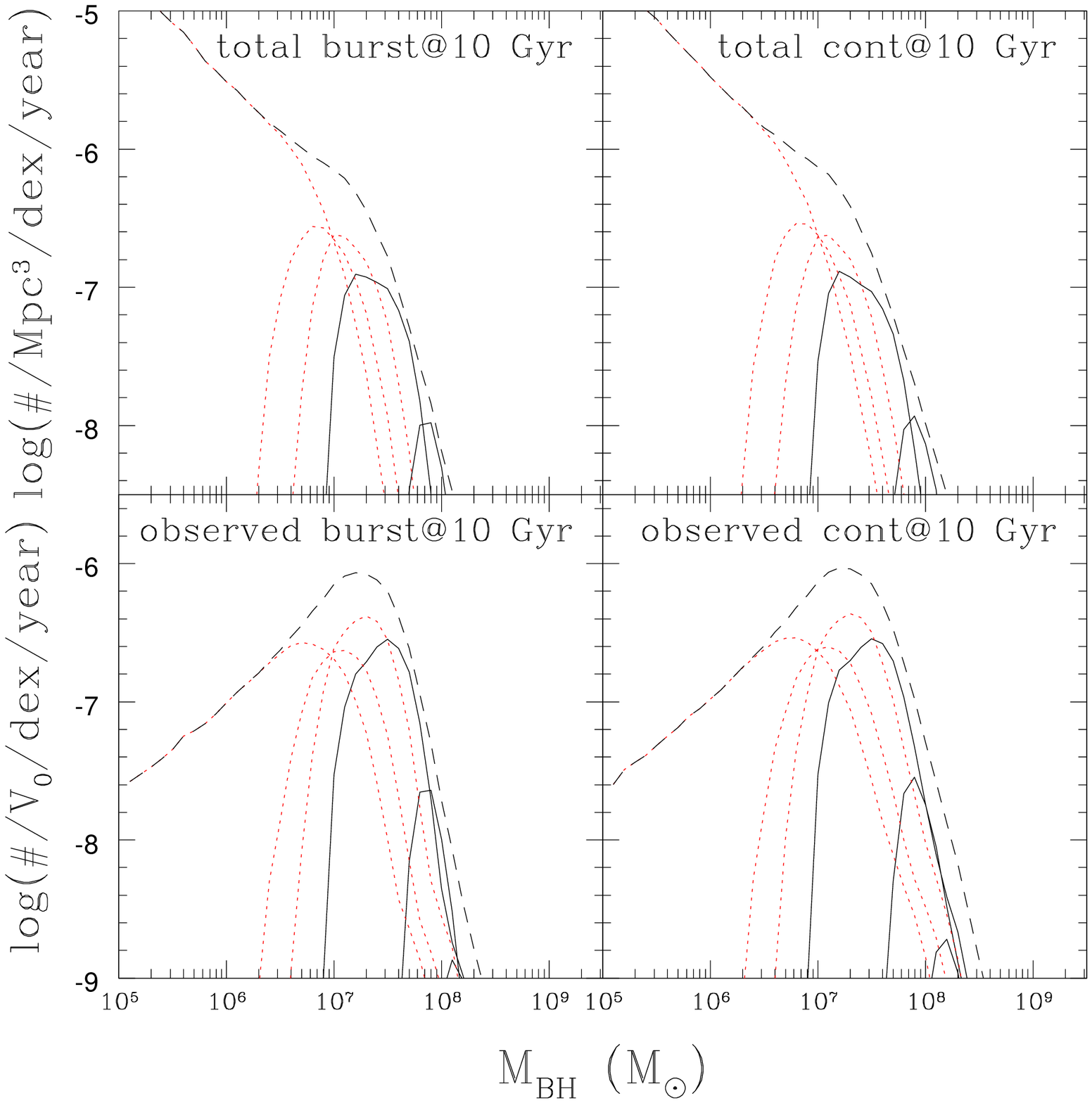}}
\caption{
  Volumetric (top) and observed (bottom) TDE rates as a function of
  the peak accretion rate in Eddington units $\dot{M}_{peak}/\dot{M}_E$
  for pinhole encounters with $n=2$ so that the peak accretion rate
  increases with the depth of the encounter as
  $\dot{M}_{peak}/\dot{M}_E \propto \beta^3$ (see Equation~\ref{eqn:mdot}).
  This Figure should be compared to Figure~\ref{fig:mdot}.
  In order of increasing accretion rate and black hole mass, the solid curves are accretion
  rates of $\dot{M}_{peak}/\dot{M}_E<10^{-1.5}$, $10^{-1.5}$-$10^{-1.0}$,
  $10^{-1.0}$-$10^{-0.5}$, and $10^{-0.5}$-$10^{0.0}$ while the
  dotted red curves are for accretion rates of $10^{0.0}$-$10^{0.5}$,
  $10^{0.5}$-$10^{1.0}$ and $>10^{1.0}$.
  The dashed curves show the total rate.  In the lower panels
  the rate scales with the fiducial volume $V_0$ (in Mpc$^3$, see text).
  }
\label{fig:mdot2}
\end{figure}

\begin{figure}
\centerline{\includegraphics[width=3.5in]{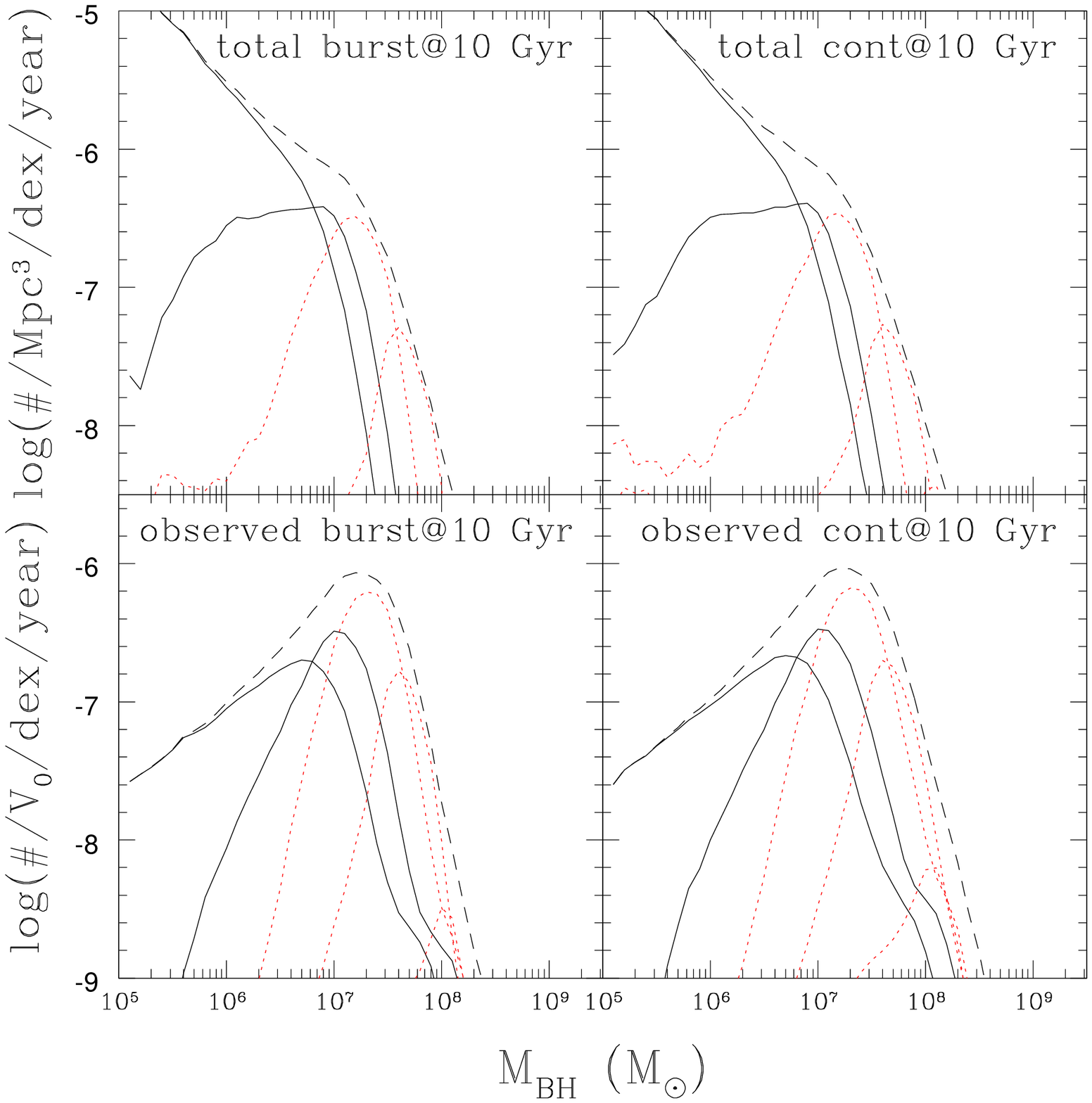}}
\caption{
  Volumetric (top) and observed (bottom) TDE rates as a function of
  the fall back time $t_{fb}$ for pinhole encounters with $n=2$ so
  that the fall back time decreases with the depth of the encounter
  as $t_{fb} \propto \beta^{-3}$ (see Equation~\ref{eqn:tfb}.
  This Figure should be compared to Figure~\ref{fig:tfb}.
  In order of increasing accretion rate and (generally) black hole mass, 
   the solid curves are for
  $t_{fb}<10^{-1.5}$~years and $10^{-1.5}$-$10^{-1.0}$,
  while the dotted red curves are for 
  $t_{fb}=10^{-1.0}$-$10^{-0.5}$, $10^{-0.5}$-$10^{0.0}$, $10^{0.0}$-$10^{0.5}$,
  $10^{0.5}$-$10^{1.0}$ and $>10^{1.0}$~years.  Events with
  the time scales of the solid curves are very likely to trigger
  present day transient searches, while the time scales
  corresponding to the dotted red curves are increasingly likely to be ignored.
  Note the contribution of long time scale events for low mass
  black holes due to the disruption of evolved stars.
  The dashed curves show the total rate.  In the lower panels
  the rate scales with the fiducial volume $V_0$ (in Mpc$^3$, see text).
  }
\label{fig:tfb2}
\end{figure}

As discussed in \S2, if the energy spread produced during an encounter is relatively
independent of the pericenter ($n=0$), then the depth of the encounter 
$\beta = R_T/R_p$ becomes a secondary variable because the basic time 
time and accretion scales are independent of $\beta$.  Figures~\ref{fig:mdot2}
and \ref{fig:tfb2} show how the distributions of pinhole events in
$\dot{M}/\dot{M}_E$ and $t_{fb}$ change
compared to Figures~\ref{fig:mdot} and \ref{fig:tfb} if we use $n=2$
for Equations~\ref{eqn:mdot} and \ref{eqn:tfb} instead of $n=0$.  The
peak accretion rates and fall back times now increase as $\beta^3$     
and decrease as $\beta^{-3}$ respectively, where $\beta=R_T/R_p$ 
characterizes the depth of the pericentric radius $R_p$ compared to
the tidal radius $R_T$ (Equation~\ref{eqn:disruption}).  The range
is limited to $R_T/R_S < \beta < 1$ with the probability distribution
of Equation~\ref{eqn:bdist}.  The overall distribution would then 
have to average over the fraction of pinhole and diffusive events
(Equation~\ref{eqn:fpin}).  

Very roughly speaking, the consequence of using $n=2$ is that the 
typical accretion rate at fixed $M_{BH}$ increases by roughly
$0.5$~dex and the typical fall back time decreases by roughly
$0.5$~dex.  The accretion rate at the peak of the expected observed
rate distribution is now mildly super-Eddington, and the typical
fall back time scale is closer to one month than three months.
These shifts would exacerbate many of the tensions in the rates
and black hole mass distributions, particularly since it makes it
more difficult to argue that some of the differences can be 
explained by transient surveys tending to ignore longer time 
scale variability at the centers of galaxies.

\end{document}